\documentclass[aps,prl,nopacs,twocolumn,preprintnumbers,notitlepage,amsmath,amssymb,superscriptaddress]{revtex4-2}
\usepackage{amssymb}
\usepackage{amsmath}
\usepackage{color}
\usepackage[pdftex]{graphicx}
\usepackage[caption=false]{subfig}
\usepackage{mathtools}
\usepackage{makecell}
\usepackage{enumerate}
\usepackage{mathtools}
\usepackage{stmaryrd}
\usepackage{enumitem}
\usepackage{textcomp}
\usepackage{gensymb}
\usepackage{notes2bib}

\newcommand{\kB}[0]{k_{\mathrm{B}}}

\begin{document}

\title{Synchronization and enhanced catalysis of mechanically coupled enzymes}

\author{Jaime Agudo-Canalejo}
\affiliation{Department of Living Matter Physics, Max Planck Institute for Dynamics and Self-Organization, D-37077 G\"ottingen, Germany}

\author{Tunrayo Adeleke-Larodo}
\affiliation{Rudolf Peierls Centre for Theoretical Physics, University of Oxford, Oxford OX1 3PU, United Kingdom}

\author{Pierre Illien}
\affiliation{Sorbonne Universit\'e, CNRS, Laboratoire Physicochimie des Electrolytes et Nanosyst\`emes Interfaciaux (PHENIX), UMR  8234, 4 place Jussieu, 75005 Paris, France}

\author{Ramin Golestanian}
\affiliation{Department of Living Matter Physics, Max Planck Institute for Dynamics and Self-Organization, D-37077 G\"ottingen, Germany}
\affiliation{Rudolf Peierls Centre for Theoretical Physics, University of Oxford, Oxford OX1 3PU, United Kingdom}

\date{\today}

\begin{abstract}

We examine the stochastic dynamics of two enzymes that are mechanically coupled to each other e.g.~through an elastic substrate or a fluid medium. The enzymes undergo conformational changes during their catalytic cycle, which itself is driven by stochastic steps along a biased chemical free energy landscape. We find conditions under which the enzymes can synchronize their catalytic steps, and discover that the coupling can lead to a significant enhancement in their overall catalytic rate. Both effects can be understood as arising from a global bifurcation in the underlying dynamical system at sufficiently strong coupling. Our findings suggest that, despite their molecular scale, enzymes can be cooperative and improve their performance in metabolic clusters.

\end{abstract}

\maketitle

\textit{Introduction.}---Since the observation of “the sympathy of two clocks” by Christiaan Huygens in 1665, synchronization phenomena have been observed in a variety of systems, at different time- and length-scales \cite{Kuramoto1984,Pikovsky2001,strogatz2012sync}. Generic theoretical frameworks, in particular the Kuramoto model, have been widely used to predict the conditions under which synchronization can occur \cite{Acebron2005}. However, the individual oscillators are usually coupled to each other through a physical medium, and it is often necessary to include the microscopic details of the mechanical coupling between them in order to obtain a useful description.
One well-studied example is the synchronization of periodically beating flagella and cilia \cite{Lauga2008,Golestanian2011}, for which hydrodynamic interactions have been shown to play a crucial role \cite{Vilfan2006,Uchida2010,Uchida2011,Uchida2011a,Uchida2012,B2015}. Another example is synchronization mediated by elastic stresses in a solid substrate, known to be important for algal flagella \cite{quaranta2015hydrodynamics,wan2016coordinated} and cardiac cells \cite{nitsan2016mechanical,cohen2016elastic}. In these micro-scale examples, the cyclic motion of each individual oscillator is driven by a non-vanishing, deterministic driving force. 

At the even smaller nanoscale, however, molecular motors and enzymes convert chemical energy into mechanical work following repeated thermodynamic cycles \cite{Magnasco1994,Julicher1997,Prost1994,Golubeva2012,Malgaretti2012}. These processes take place in a noise-activated regime, where motion only occurs stochastically, in response to barrier-crossing events along a chemical free energy landscape. Recently, the relationship between the conformational changes of enzymes during their catalytic cycle \cite{glowacki2012taking,callender2015dynamical} and their translational dynamics has been the subject of many experimental and theoretical studies \cite{Gunther2018a,Ross2019,RG2010,Golestanian2015, Mikhailov2015,bai2015hydrodynamics,Illien2017b,Agudo-Canalejo2018a,hosaka2020mechanochemical,koyano2020diffusion}. While these studies have considered the effect of enzymatic activity on mechanical motion, how mechanical interactions feed back into enzymatic activity and whether synchronization of the catalytic cycles across enzymes is possible are questions that remain open. These questions are of high biological relevance considering that enzymes are frequently assembled into clusters \cite{sweetlove2018role,kerfeld2018bacterial,liu2016cytoophidium,selwood2012dynamic,gomperts1996clustering,shuai2003optimal}.

In this Letter, we study the dynamics of two enzymes which undergo conformational changes during their catalytic cycle and interact with each other mechanically. We show that this coupling is sufficient to synchronize their stochastic catalytic steps, and moreover leads to a significant enhancement of their catalytic rate.

\begin{figure}[b]
	\begin{center}
		\includegraphics[width=1\linewidth]{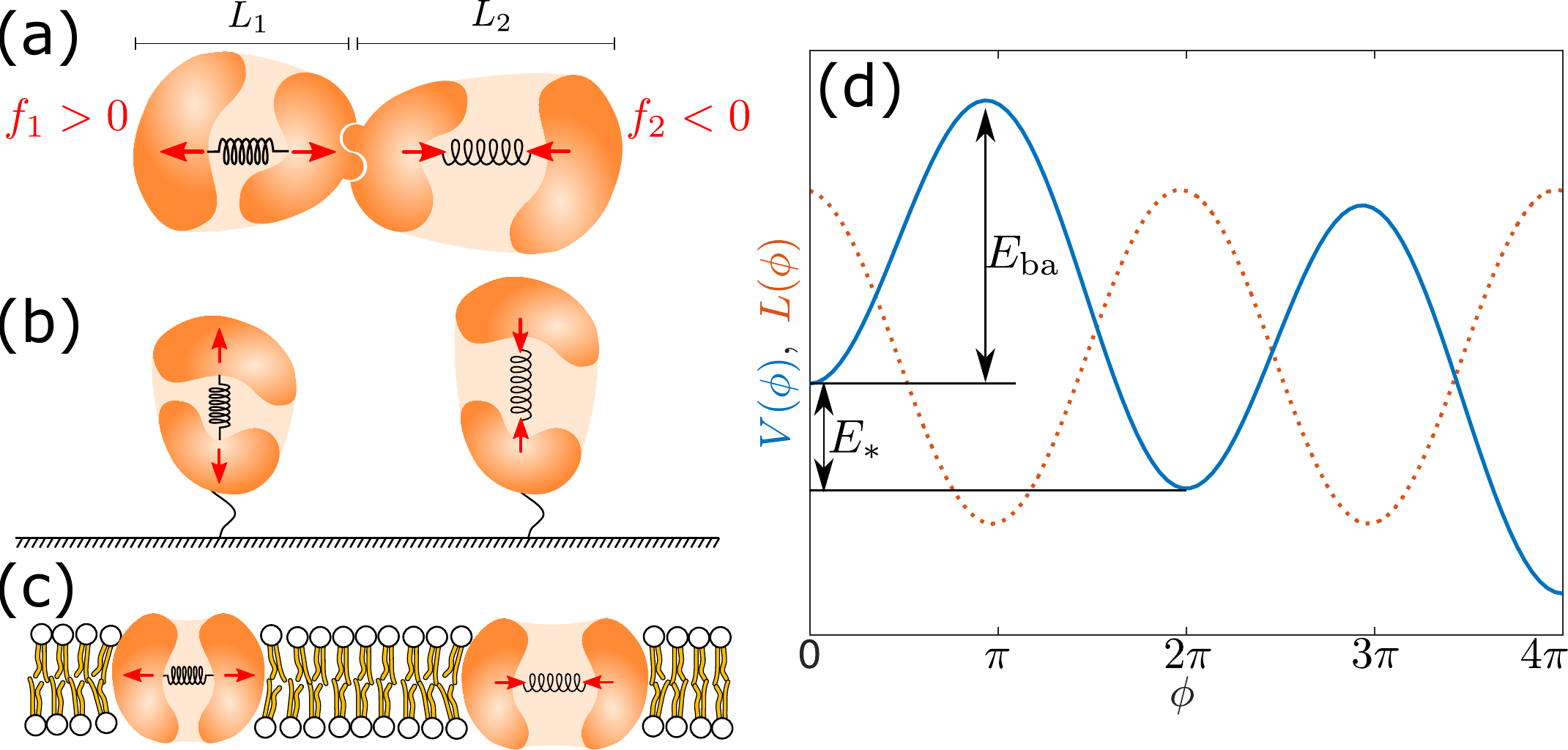}
		\caption{Examples of mechanical interactions: (a) Two enzymes bound to each other forming a complex. Each enzyme has elongation $L_\alpha$ and experiences an internal force $f_\alpha$. (b) Two enzymes interact with each other hydrodynamically through the surrounding viscous fluid medium. (c) Two enzymes embedded in a lipid membrane interact elastically. (d) The catalytic cycle of each enzyme is represented by a phase $\phi_\alpha$ evolving in a biased free energy landscape $V(\phi_\alpha)$ (solid black). The enzyme elongation $L_\alpha$ tries to adapt to a phase-dependent rest length $L(\phi_\alpha)$ (dotted red).  \label{fig:coupling}}
	\end{center}
\end{figure}

\textit{Mechanochemical coupling.}---We use a minimal model where each enzyme $\alpha$ is considered to have a single mechanical degree of freedom $L_\alpha$, which might represent for example its elongation,
see Fig.~\ref{fig:coupling}(a--c); and a single chemical degree of freedom or phase $\phi_\alpha$, which is a reaction coordinate describing the state of the chemical reaction happening inside the enzyme. Both $L_\alpha$ and $\phi_\alpha$ evolve together according to the potential $U(L_\alpha,\phi_\alpha) = \frac{k}{2} [L_\alpha - L(\phi_\alpha)]^2 + V(\phi_\alpha)$. Here, the first term describes conformational changes of the enzyme during the catalytic cycle, with $L(\phi_\alpha)$ being the rest length of the enzyme as a function of the reaction coordinate and $k$ the stiffness of the enzyme, while the second term represents the free energy of the reaction, described by a biased potential $V(\phi_\alpha)$ that drives the phase forward \cite{Magnasco1994,Julicher1997}; see Fig.~\ref{fig:coupling}(d). Assuming that the overdamped medium surrounding the enzymes couples forces to velocities linearly, the dynamics of the elongation of enzyme $\alpha$ will be governed by $\dot{L}_\alpha=\mu \big(f_\alpha + h f_\beta \big)$, where $\mu$ is the mobility associated to the elongation, and $h$ is the (dimensionless) mechanical coupling between the two enzymes $\alpha$ and $\beta \neq \alpha$. As an example, for enzymes directly coupled into a complex as in Fig.~\ref{fig:coupling}(a) the coupling constant $h$ can be easily calculated and is found to be negative with $0>h>-1$ \cite{suppmat}. The internal forces (force-dipoles) $f_\alpha$ and $f_\beta$ generated by the corresponding enzymes can  be calculated as $f_\alpha =- \partial_{L_\alpha} U(L_\alpha,\phi_\alpha) = - k [L_\alpha - L(\phi_\alpha)]$.  In turn, the phase dynamics are given by $\dot{\phi}_\alpha = -\mu_\phi \partial_{\phi_\alpha} U(L_\alpha,\phi_\alpha) = -\mu_\phi \left\{ - k [L_\alpha - L(\phi_\alpha)] L'(\phi_\alpha) + V'(\phi_\alpha)  \right\}$, where $\mu_\phi$ is the mobility along the chemical reaction coordinate.

\begin{figure*}[t]
	\begin{center}
		\includegraphics[width=1\linewidth]{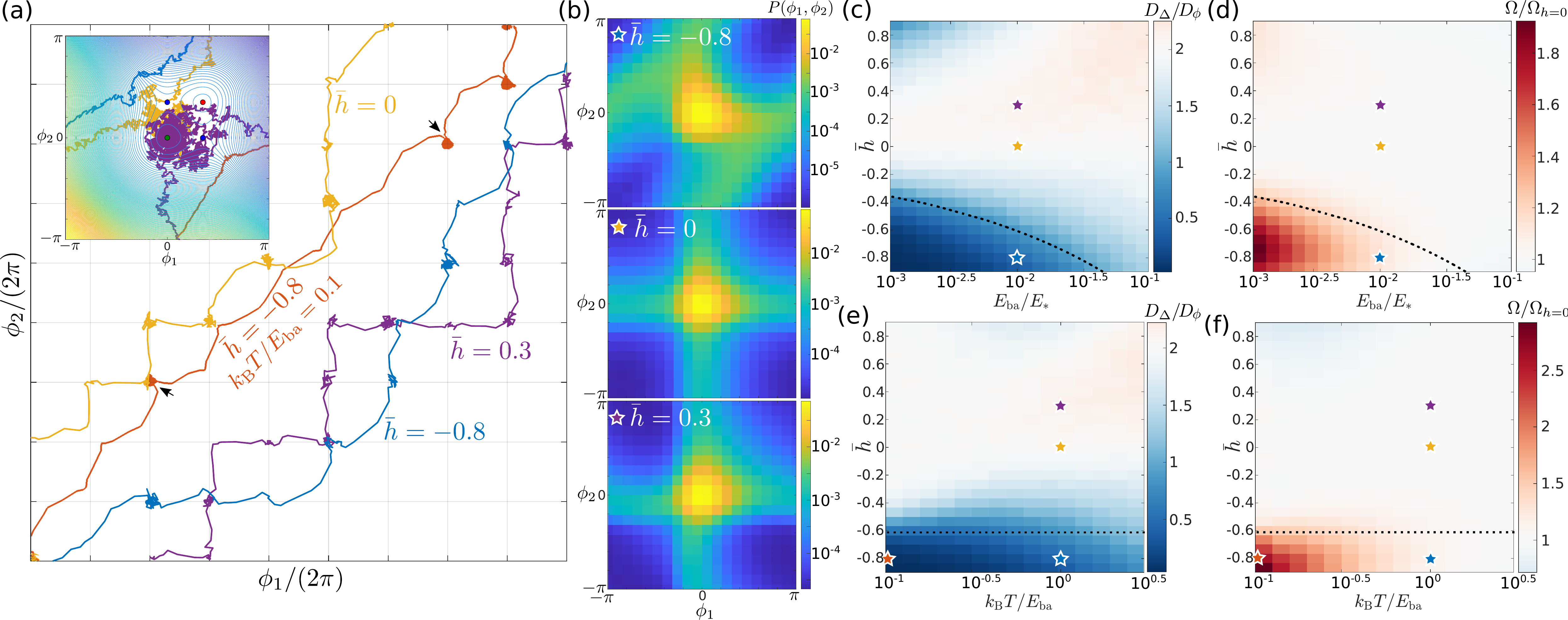}
		\caption{(a) Trajectories of the system in $(\phi_1,\phi_2)$ space for different values of $\bar{h}$, with $E_\mathrm{ba}/E_*=10^{-2}$ and $k_\mathrm{B}T/E_\mathrm{ba}=1$ (except where noted). The grey grid marks integer values of $\phi_{1,2}/(2\pi)$. Trajectories for zero or positive coupling show mostly horizontal and vertical segments, implying single-enzyme steps, whereas those for negative coupling ($\bar{h}=-0.8$) are diagonal, implying synchronized two-enzyme steps. For $\bar{h}=-0.8$ and low noise, multi-step diagonal runs are observed (red line, between the two arrows). The inset shows magnified segments of the same trajectories (same colors) overlayed onto a contour map of the chemical free energy landscape $V(\phi_1)+V(\phi_2)$. The green circle, red circle, and two black circles represent the stable fixed point, unstable fixed point, and two saddle points of the landscape, respectively. Notice how the trajectories for negative coupling avoid the stable fixed point, while those for zero or positive coupling do spend a significant amount of time around it. (b) Probability distribution $P(\phi_1,\phi_2)$, with phases modulo $2\pi$, for the same trajectories as in (a). Note the diagonal stripes for  $\bar{h}=-0.8$. (c--f) Heatmaps for the phase-difference diffusion coefficient $D_\Delta$ (c,e) and the catalytic rate $\Omega$ (d,f), as a function of $E_\mathrm{ba}/E_*$ and $\bar{h}$ for fixed $k_\mathrm{B}T/E_\mathrm{ba}=1$ (c,d), and of $k_\mathrm{B}T/E_\mathrm{ba}$ and $\bar{h}$ for fixed $E_\mathrm{ba}/E_*=10^{-2}$ (e,f). The dashed lines in (c--f) correspond to the synchronization boundary based on the deterministic phase portraits; see Fig.~\ref{fig:phaseportraits}.  \label{fig:simulations}}
	\end{center}
\end{figure*}

\textit{Phase equations.}---The coupled dynamics of length and phase can be simplified further by projecting the dynamics of the lengths, assumed fast, onto the slow manifold of the configuration space defined by the phases $\phi_\alpha$ \cite{suppmat}. Staying to lowest order in $A_\alpha \equiv \sqrt{\frac{\mu_\phi}{\mu}} L'(\phi_\alpha)$, which corresponds to the assumption that phase changes rather than conformational changes constitute the bottleneck in the dynamics of the enzyme \cite{nonzeromu}, the  deterministic dynamics for the phases reads
\begin{equation}
\dot{\phi}_{\alpha}(t)=\omega (\phi_{\alpha})+ \frac{ h A_\alpha A_\beta}{1-h^2} \omega(\phi_{\beta})
\label{eq:phasedynamics}
\end{equation}
with $\omega(\phi_\alpha) \equiv -\mu_\phi  V'(\phi_\alpha)$.
This shows that the reaction dynamics of two enzymes that undergo conformational changes during catalysis are coupled through the mechanical interaction. We note that, in contrast to usual models of synchronization, the interaction term is proportional to the driving force $\omega(\phi_\alpha)$. In fact, whereas in other models the coupling term can be understood as coming from the gradient of a potential (e.g.~a term proportional to $-\cos(\phi_\alpha-\phi_\beta)$ in the Kuramoto model), such a description is not possible here: the velocity field corresponding to the right hand side of Eq.~\eqref{eq:phasedynamics} has non-zero curl when $h\neq 0$. 
Moreover, while in typical descriptions of synchronization the driving force $\omega(\phi_\alpha)$ is either a constant or a positive-definite function of the phase that can be mapped onto a constant using a gauge transformation \cite{Acebron2005}, in our system $\omega(\phi_\alpha)$ vanishes and changes sign twice through a complete catalytic cycle, because of the energy barrier in $V(\phi)$. This highlights that catalysis is an activated process that can only occur in the presence of noise.

\textit{Stochastic dynamics.}---Thermal noise can be systematically added to the deterministic dynamics in Eq.~(\ref{eq:phasedynamics}) \cite{suppmat}, resulting in the stochastic dynamics
\begin{eqnarray}
&&\dot{\phi}_\alpha(t)=M_{\alpha \beta}\left[-\mu_\phi V'(\phi_\beta)\right]+k_\mathrm{B}T\mu_\phi \, \Sigma_{\alpha \nu} \partial_\beta \Sigma_{\beta \nu} \nonumber \\
&&\hskip1cm +\sqrt{2 k_\mathrm{B}T \mu_\phi} \,\Sigma_{\alpha \beta} \,\xi_\beta(t) ,
\label{eq:phasedynamics-noisy}
\end{eqnarray}
where the mobility tensor is defined as $M_{11}=M_{22}=1$, and $M_{12}=M_{21}= h A_1 A_2 / (1-h^2)$, the square-root of the mobility tensor $\Sigma$ is defined via $M_{\alpha \beta}=\Sigma_{\alpha \nu} \Sigma_{\beta \nu}$ \cite{suppmat}, and Einstein summation convention for repeated indices is used. Eq.~(\ref{eq:phasedynamics-noisy}) is to be interpreted in the Stratonovich sense. The first term is identical to the deterministic dynamics in Eq.~(\ref{eq:phasedynamics}), the last term is a (multiplicative) noise where $\xi$ satisfies $\langle \xi_\alpha(t) \xi_\beta(t') \rangle = \delta_{\alpha \beta} \delta(t-t')$, and the second term is the spurious drift associated with this multiplicative noise. These dynamics are constructed such that they correspond to the Fokker-Planck equation $\partial_t {\cal P} = \partial_{\alpha} \left[M_{\alpha \beta} \mu_\phi \Big( V'(\phi_\beta) {\cal P}+k_\mathrm{B}T  \partial_{\beta}{\cal P}\Big)\right]$
for the probability distribution ${\cal P}(\phi_1,\phi_2;t)$, which ensures equilibration to the Boltzmann distribution ${\cal{P}}(\phi_1,\phi_2)\propto \exp\{-{[V(\phi_1)+V(\phi_2)]}/{\kB T}\}$, independently of the value of $h$, whenever the system allows equilibration, e.g.~when the potential $V(\phi)$ is unbiased or the range of $\phi_1,\phi_2$ is bounded.
Eq.~\eqref{eq:phasedynamics-noisy} highlights that the forces on the phases are actually conservative, with associated potential $V(\phi_1)+V(\phi_2)$, but are connected to the phase velocities \emph{via} a non-diagonal mobility matrix, with the coupling constant $h$ setting the magnitude of the off-diagonal terms, which become important when the system is out of equilibrium. As a side-note, we believe that these physically-suggestive properties make Eq.~\eqref{eq:phasedynamics-noisy} stand out in the context of synchronization (as an unexplored class of dynamical systems), not just in the particular form of the off-diagonal terms that we have derived here from mechanical coupling, but in its full generality  \cite{constantcoupling}.

\textit{Brownian dynamics simulations.}---In what follows, we model the reaction free energy using a washboard potential $V(\phi) = - F \phi - v \cos \left[\phi + \arcsin(F/v)\right]$, which has minima at $\phi=2\pi n$ for all $n \in \mathbb{Z}$ when $F/v<1$. Here, $F$ and $v$ determine the height of the free energy barrier $E_\mathrm{ba}$ and the free energy difference of the chemical reaction $E_*$ [see Fig.~\ref{fig:coupling}(c)] through $E_\mathrm{ba} = \left[ 2 \sqrt{1-(F/v)^2}-(F/v)(\pi-2\arcsin(F/v)) \right] v$ and $E_* = 2\pi F$. The washboard potential thus represents an unending sequence of substrate-to-product transformation reactions, assuming that substrate is abundant and that the binding and unbinding of substrate and product in between each reaction happen very fast compared to all other processes. The conformational changes of the enzyme mimic the washboard potential, with the rest length given by $L(\phi) = L_0 + \ell \cos \left[\phi + \arcsin(F/v)\right]$, so that the extrema of $V'(\phi)$ coincide with those of $L'(\phi)$. Here, $\ell$ represents the amplitude of the conformational changes and may be positive or negative depending on whether the enzyme expands or contracts during catalysis. The synchronization dynamics, however, is independent of the sign of $\ell$, as the equations of motion are invariant under changes of this sign.
 Defining dimensionless time as $\tau \equiv t \mu_\phi v$, the system depends only on three dimensionless parameters: the rescaled coupling constant $\bar{h} \equiv h \mu_\phi \ell^2 / [\mu (1-h^2)]$, the bias of the free energy landscape as determined by $E_\mathrm{ba}/E_*$, and the noise strength $k_\mathrm{B}T/E_\mathrm{ba}$.

Results of Brownian dynamics simulations of Eq.~(\ref{eq:phasedynamics-noisy}) are shown in Fig.~\ref{fig:simulations} and Fig.~S1 in the Supplemental Material \cite{suppmat}. As expected, the enzymes undergo stochastic, quasi-discrete steps. While for low or positive coupling $\bar{h}$ the two enzymes mostly do individual catalytic steps, for sufficiently negative $\bar{h}$ the two enzymes tend to step in synchrony; see Fig.~\ref{fig:simulations}(a,b). Remarkably, for sufficiently low noise ($k_\mathrm{B}T/E_\mathrm{ba}\ll 1$), we observe long synchronized runs, in which the two enzymes undergo a large number of joint catalytic steps after a thermal fluctuation kicks them out of a local free energy minimum (and before falling back into a minimum). An example of a 5-step run can be seen between the two black arrows for the red trajectory in Fig.~\ref{fig:simulations}(a).

As a quantitative measure of synchronization, we use the phase-difference diffusion coefficient $D_\Delta$, calculated from the relation $\langle (\phi_1-\phi_2)^2 \rangle \sim 2 D_\Delta t$. This can be compared to the single-phase diffusion coefficient $D_\phi$ calculated from $\langle (\phi_\alpha- \langle \phi_\alpha \rangle)^2 \rangle  \sim 2 D_\phi t$. If $\phi_1$ and $\phi_2$ were independent variables, we would expect $D_\Delta/D_\phi=2$, as we observe for $\bar{h}=0$. However, negative values of $\bar{h}$ lead to $D_\Delta/D_\phi<2$, implying that synchronous steps are favored; see Fig.~\ref{fig:simulations}(c,e). Synchronization is most pronounced for strong bias ($E_\mathrm{ba}/E_* \ll 1$) and low noise ($k_\mathrm{B}T/E_\mathrm{ba}\ll 1$). For positive $\bar{h}$ we find
predominantly $D_\Delta/D_\phi \gtrsim 2$, implying in this case that synchronous steps are inhibited. We then consider the total catalytic activity, i.e.~the number of catalytic steps per enzyme per unit time over the whole time of the simulation $\Omega \equiv [\phi_1(\tau_\mathrm{tot})+\phi_2(\tau_\mathrm{tot})]/(4\pi \tau_\mathrm{tot})$; see Fig.~\ref{fig:simulations}(d,f). We find that mechanical coupling can enhance catalytic activity, particularly for strongly synchronized cases (with strong bias, low noise, and negative $\bar{h}$), in which case enhancements as large as 200\% are seen, a remarkable observation for just two coupled enzymes.

\begin{figure*}[t!]
	\begin{center}
		\includegraphics[width=1\linewidth]{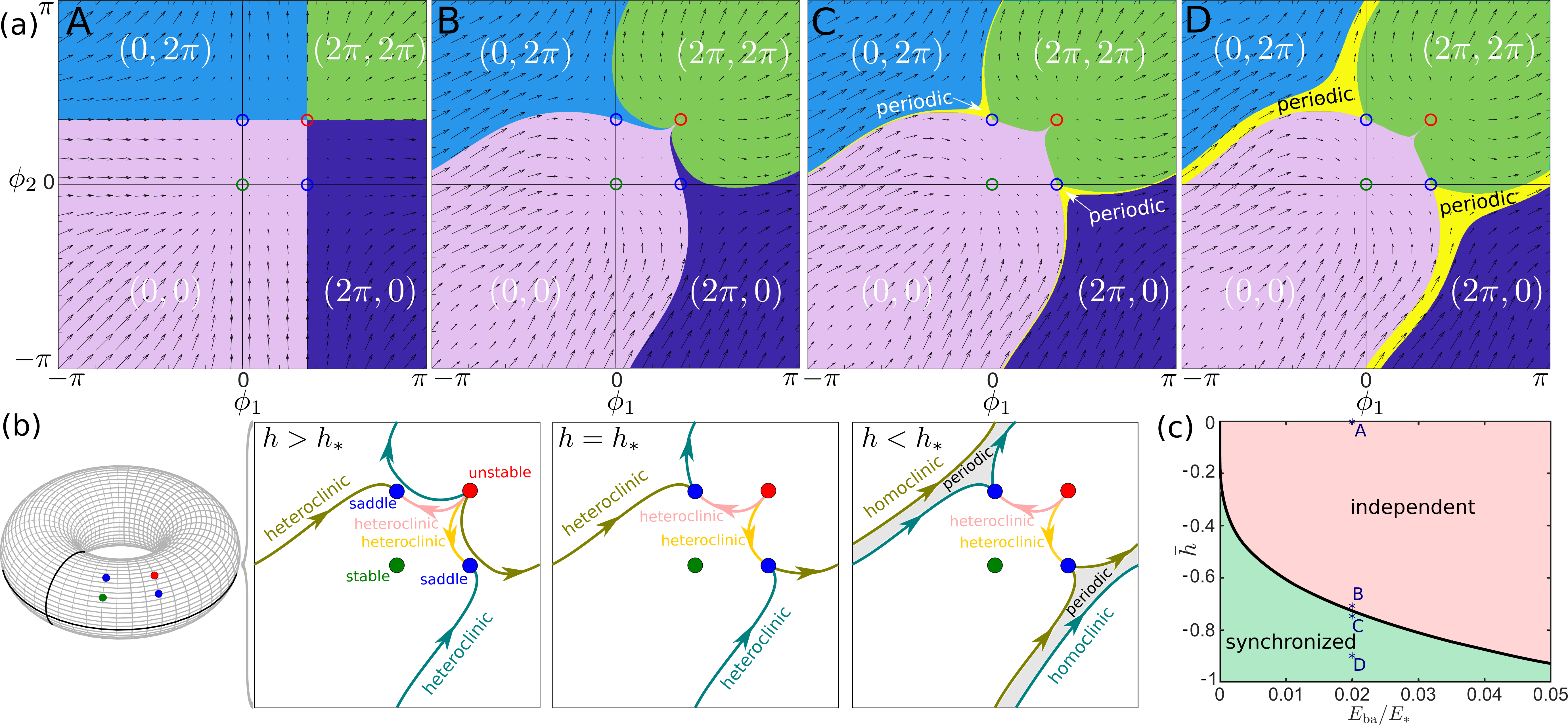}
		\caption{(a) Phase portraits for $E_\mathrm{ba}/E_*=0.02$ and $\bar{h}=0, -0.71,-0.75,-0.9$, corresponding to the points A--D in the phase diagram in (c). Independently of the value of $\bar{h}$, the system always has a stable fixed point at $(0,0)$ [green circle], an unstable fixed point at $(\phi_*,\phi_*)$ where $\phi_*\equiv \pi - 2 \arcsin{(F/v)}$ corresponds to the location of the maximum of $V(\phi)$ [red circle], and two saddle points at $(\phi_*,0)$ and $(0,\phi_*)$ [blue circles]. The colored regions labeled $(0,0)$, $(0,2\pi)$, $(2\pi,0)$, and $(2\pi,2\pi)$ represent four different basins of attraction, all of which go to the stable fixed point $(0,0)$, but with trajectories that wind differently around the torus on the way there. Notice the topological transition occurring between B and C, where a \emph{running band} of periodic orbits emerges (in yellow). (b) The  transition can be understood as a global bifurcation of the dynamical system on the torus with decreasing $h$. At $h=h_*$, two pairs of heteroclinic orbits connecting the unstable point to the saddle points collide. For $h<h_*$, two homoclinic orbits arise, with the running band between them. (c) Phase diagram based on the topology of the deterministic phase portraits.   \label{fig:phaseportraits}}
	\end{center}
\end{figure*}

\textit{Phase portrait.}---The fact that synchronization and enhanced catalysis are strongest at low noise suggests that their emergence may be understood from the  phase portraits of the underlying deterministic dynamical system, Eq.~(\ref{eq:phasedynamics}). Indeed, the dynamical system, which is defined on the torus, undergoes a global bifurcation for sufficiently strong negative coupling $\bar{h}$; see Fig.~\ref{fig:phaseportraits}(a,b). For $\bar{h}>\bar{h}_*$, the phase space is divided into four basins of attraction, separated by four heteroclinic orbits. At a critical value $\bar{h}=\bar{h}_*$, the heteroclinic orbits change topology, and for $\bar{h}<\bar{h}_*$ two of the heteroclinic orbits become two homoclinic orbits, between which a {\em running band} of periodic orbits emerges. 

This topological bifurcation has important consequences. Whereas for $\bar{h}>\bar{h}_*$ thermal fluctuations will typically kick a system which is initially at $(0,0)$ into either of the basins of attraction $(2\pi,0)$ or $(0,2\pi)$ corresponding to a single-enzyme step, for $\bar{h} < \bar{h}_*$ the system is instead kicked into either the basin of attraction $(2\pi,2\pi)$ corresponding to a synchronized two-enzyme step, or into the running band. In the latter case, the system will perform multiple synchronized steps until noise kicks it out of the band once again. The critical value $\bar{h}_*$ decreases with increasing $E_\mathrm{ba}/E_*$, as seen in Fig.~\ref{fig:phaseportraits}(c), and appears to be well described by $\bar{h}_* \simeq - 2 (E_\mathrm{ba}/E_*)^{1/4}$. This value is plotted as the dashed line in Fig.~\ref{fig:simulations}(c--f), and correctly predicts the regions with enhanced catalysis and synchronization.

\textit{Discussion.}---Using a minimal model, we have shown that enzymes that undergo conformational changes during their catalytic cycle can synchronize with each other through mechanical interactions, which moreover can significantly enhance their overall catalytic rate. These effects are favored for negative mechanical coupling $h<0$, which implies that contraction of one enzyme favors expansion of the other, and vice versa. A negative coupling is guaranteed for complexed enzymes as in Fig.~\ref{fig:coupling}(a) \cite{suppmat}, and should be expected in similar configurations such as the one in Fig.~\ref{fig:coupling}(c). Here, synchronization arises as an entrainment of the inherently stochastic, noise-activated catalytic steps of the two enzymes. While synchronization in excitable systems has been described before \cite{lindner2004effects}, particularly in the context of FitzHugh-Nagumo oscillators \cite{neiman1999noise,hu2000phase,sosnovtseva2001clustering}, the coupling in these systems was by means of an added coupling force (diffusive or Kuramoto-like) and resulted in a synchronization transition mediated by standard Hopf or saddle-node bifurcations. In contrast, in our system we find a novel form of coupling which arises from off-diagonal terms in the mobility matrix that connects forces to velocities, and thus leaves the equilibrium probability distribution of the system intact while introducing non-trivial effects in an out-of-equilibrium setting. The resulting transition is mediated by a global bifurcation which gives not only synchronization but also enhanced catalysis. We note that the mechanism for enhanced catalysis that we observe is also distinct from recent proposals for activated barrier crossing \cite{PhysRevLett.122.258001,PhysRevLett.124.118002,woillez2020nonlocal} 
which rely on colored noise.

Biological enzymes often form dense assemblies where mechanical interactions should be expected. These range from large-scale, three-dimensional clusters such as ``metabolons'' \cite{sweetlove2018role} and bacterial micro-compartments \cite{kerfeld2018bacterial}, to ordered filaments such as the cytoophidium \cite{liu2016cytoophidium}, to oligomeric complexes of just a few enzymes \cite{selwood2012dynamic}. While the functional benefit of these structures is not yet clear, it has been proposed that proximity favors channeling of reaction intermediates among different enzymes in the same catalytic pathway \cite{sweetlove2018role}. Our result of enhanced catalysis provides a possible additional advantage to close proximity between enzymes. In fact, many enzymes that are functional in their monomeric form but also assemble into homooligomeric forms [see Fig.~\ref{fig:coupling}(a)] are more catalytically active in the oligomeric form \cite{selwood2012dynamic}, a behavior which could be explained by mechanical coupling as proposed here. Moreover, synchronization effects could be particularly relevant in the context of rapid and robust signaling by membrane ion channels, which also operate in clusters \cite{gomperts1996clustering,shuai2003optimal}. Future experiments may also test our predictions in a controlled \emph{in vitro} setting, by creating enzyme assemblies with designed geometry \cite{dueber2009synthetic,kufer2008single,wilner2009enzyme}, or using single-molecule techniques that allow for the measurement of individual catalytic events and conformational changes \cite{min2005fluctuating,gumpp2009triggering,choi2012single,pelz2016subnanometre}.

This work has received support from the Max Planck School Matter to Life and the MaxSynBio Consortium, which are jointly funded by the Federal Ministry of Education and Research (BMBF) of Germany, and the Max Planck Society.
T.A.-L.~acknowledges the support of an EPSRC Studentship.

\bibliography{bib_synchronisation.bib}

\begin{thebibliography}{57}%
\makeatletter
\providecommand \@ifxundefined [1]{%
 \@ifx{#1\undefined}
}%
\providecommand \@ifnum [1]{%
 \ifnum #1\expandafter \@firstoftwo
 \else \expandafter \@secondoftwo
 \fi
}%
\providecommand \@ifx [1]{%
 \ifx #1\expandafter \@firstoftwo
 \else \expandafter \@secondoftwo
 \fi
}%
\providecommand \natexlab [1]{#1}%
\providecommand \enquote  [1]{``#1''}%
\providecommand \bibnamefont  [1]{#1}%
\providecommand \bibfnamefont [1]{#1}%
\providecommand \citenamefont [1]{#1}%
\providecommand \href@noop [0]{\@secondoftwo}%
\providecommand \href [0]{\begingroup \@sanitize@url \@href}%
\providecommand \@href[1]{\@@startlink{#1}\@@href}%
\providecommand \@@href[1]{\endgroup#1\@@endlink}%
\providecommand \@sanitize@url [0]{\catcode `\\12\catcode `\$12\catcode
  `\&12\catcode `\#12\catcode `\^12\catcode `\_12\catcode `\%12\relax}%
\providecommand \@@startlink[1]{}%
\providecommand \@@endlink[0]{}%
\providecommand \url  [0]{\begingroup\@sanitize@url \@url }%
\providecommand \@url [1]{\endgroup\@href {#1}{\urlprefix }}%
\providecommand \urlprefix  [0]{URL }%
\providecommand \Eprint [0]{\href }%
\providecommand \doibase [0]{https://doi.org/}%
\providecommand \selectlanguage [0]{\@gobble}%
\providecommand \bibinfo  [0]{\@secondoftwo}%
\providecommand \bibfield  [0]{\@secondoftwo}%
\providecommand \translation [1]{[#1]}%
\providecommand \BibitemOpen [0]{}%
\providecommand \bibitemStop [0]{}%
\providecommand \bibitemNoStop [0]{.\EOS\space}%
\providecommand \EOS [0]{\spacefactor3000\relax}%
\providecommand \BibitemShut  [1]{\csname bibitem#1\endcsname}%
\let\auto@bib@innerbib\@empty
\bibitem [{\citenamefont {Kuramoto}(1984)}]{Kuramoto1984}%
  \BibitemOpen
  \bibfield  {author} {\bibinfo {author} {\bibfnamefont {Y.}~\bibnamefont
  {Kuramoto}},\ }\href {https://doi.org/10.1017/CBO9781107415324.004} {\emph
  {\bibinfo {title} {{Chemical Oscillations, Waves and Turbulence}}}}\
  (\bibinfo  {publisher} {Springer},\ \bibinfo {address} {New York},\ \bibinfo
  {year} {1984})\BibitemShut {NoStop}%
\bibitem [{\citenamefont {Pikovsky}\ \emph {et~al.}(2001)\citenamefont
  {Pikovsky}, \citenamefont {Rosenblum},\ and\ \citenamefont
  {Kurths}}]{Pikovsky2001}%
  \BibitemOpen
  \bibfield  {author} {\bibinfo {author} {\bibfnamefont {A.}~\bibnamefont
  {Pikovsky}}, \bibinfo {author} {\bibfnamefont {M.}~\bibnamefont
  {Rosenblum}},\ and\ \bibinfo {author} {\bibfnamefont {J.}~\bibnamefont
  {Kurths}},\ }\href {https://doi.org/10.1017/cbo9780511755743} {\emph
  {\bibinfo {title} {Synchronization}}}\ (\bibinfo  {publisher} {Cambridge
  University Press},\ \bibinfo {year} {2001})\BibitemShut {NoStop}%
\bibitem [{\citenamefont {Strogatz}(2012)}]{strogatz2012sync}%
  \BibitemOpen
  \bibfield  {author} {\bibinfo {author} {\bibfnamefont {S.~H.}\ \bibnamefont
  {Strogatz}},\ }\href@noop {} {\emph {\bibinfo {title} {Sync: How order
  emerges from chaos in the universe, nature, and daily life}}}\ (\bibinfo
  {publisher} {Hachette UK},\ \bibinfo {year} {2012})\BibitemShut {NoStop}%
\bibitem [{\citenamefont {Acebr{\'{o}}n}\ \emph {et~al.}(2005)\citenamefont
  {Acebr{\'{o}}n}, \citenamefont {Bonilla}, \citenamefont {Vicente},
  \citenamefont {Ritort},\ and\ \citenamefont {Spigler}}]{Acebron2005}%
  \BibitemOpen
  \bibfield  {author} {\bibinfo {author} {\bibfnamefont {J.~A.}\ \bibnamefont
  {Acebr{\'{o}}n}}, \bibinfo {author} {\bibfnamefont {L.~L.}\ \bibnamefont
  {Bonilla}}, \bibinfo {author} {\bibfnamefont {C.~J.~P.}\ \bibnamefont
  {Vicente}}, \bibinfo {author} {\bibfnamefont {F.}~\bibnamefont {Ritort}},\
  and\ \bibinfo {author} {\bibfnamefont {R.}~\bibnamefont {Spigler}},\
  }\bibfield  {title} {\bibinfo {title} {{The Kuramoto model: A simple paradigm
  for synchronization phenomena}},\ }\href
  {https://doi.org/10.1103/RevModPhys.77.137} {\bibfield  {journal} {\bibinfo
  {journal} {Rev. Mod. Phys.}\ }\textbf {\bibinfo {volume} {77}},\ \bibinfo
  {pages} {137} (\bibinfo {year} {2005})}\BibitemShut {NoStop}%
\bibitem [{\citenamefont {Lauga}\ and\ \citenamefont
  {Powers}(2009)}]{Lauga2008}%
  \BibitemOpen
  \bibfield  {author} {\bibinfo {author} {\bibfnamefont {E.}~\bibnamefont
  {Lauga}}\ and\ \bibinfo {author} {\bibfnamefont {T.~R.}\ \bibnamefont
  {Powers}},\ }\bibfield  {title} {\bibinfo {title} {{The hydrodynamics of
  swimming microorganisms}},\ }\href
  {https://doi.org/10.1088/0034-4885/72/9/096601} {\bibfield  {journal}
  {\bibinfo  {journal} {Rep. Prog. Phys.}\ }\textbf {\bibinfo {volume} {72}},\
  \bibinfo {pages} {096601} (\bibinfo {year} {2009})}\BibitemShut {NoStop}%
\bibitem [{\citenamefont {Golestanian}\ \emph {et~al.}(2011)\citenamefont
  {Golestanian}, \citenamefont {Yeomans},\ and\ \citenamefont
  {Uchida}}]{Golestanian2011}%
  \BibitemOpen
  \bibfield  {author} {\bibinfo {author} {\bibfnamefont {R.}~\bibnamefont
  {Golestanian}}, \bibinfo {author} {\bibfnamefont {J.~M.}\ \bibnamefont
  {Yeomans}},\ and\ \bibinfo {author} {\bibfnamefont {N.}~\bibnamefont
  {Uchida}},\ }\bibfield  {title} {\bibinfo {title} {Hydrodynamic
  synchronization at low reynolds number},\ }\href
  {https://doi.org/10.1039/c0sm01121e} {\bibfield  {journal} {\bibinfo
  {journal} {Soft Matter}\ }\textbf {\bibinfo {volume} {7}},\ \bibinfo {pages}
  {3074} (\bibinfo {year} {2011})}\BibitemShut {NoStop}%
\bibitem [{\citenamefont {Vilfan}\ and\ \citenamefont
  {J{\"{u}}licher}(2006)}]{Vilfan2006}%
  \BibitemOpen
  \bibfield  {author} {\bibinfo {author} {\bibfnamefont {A.}~\bibnamefont
  {Vilfan}}\ and\ \bibinfo {author} {\bibfnamefont {F.}~\bibnamefont
  {J{\"{u}}licher}},\ }\bibfield  {title} {\bibinfo {title} {{Hydrodynamic flow
  patterns and synchronization of beating cilia}},\ }\href
  {https://doi.org/10.1103/PhysRevLett.96.058102} {\bibfield  {journal}
  {\bibinfo  {journal} {Phys. Rev. Lett.}\ }\textbf {\bibinfo {volume} {96}},\
  \bibinfo {pages} {058102} (\bibinfo {year} {2006})}\BibitemShut {NoStop}%
\bibitem [{\citenamefont {Uchida}\ and\ \citenamefont
  {Golestanian}(2010)}]{Uchida2010}%
  \BibitemOpen
  \bibfield  {author} {\bibinfo {author} {\bibfnamefont {N.}~\bibnamefont
  {Uchida}}\ and\ \bibinfo {author} {\bibfnamefont {R.}~\bibnamefont
  {Golestanian}},\ }\bibfield  {title} {\bibinfo {title} {{Synchronization and
  collective dynamics in a carpet of microfluidic rotors}},\ }\href
  {https://doi.org/10.1103/PhysRevLett.104.178103} {\bibfield  {journal}
  {\bibinfo  {journal} {Phys. Rev. Lett.}\ }\textbf {\bibinfo {volume} {104}},\
  \bibinfo {pages} {178103} (\bibinfo {year} {2010})}\BibitemShut {NoStop}%
\bibitem [{\citenamefont {Uchida}\ and\ \citenamefont
  {Golestanian}(2011)}]{Uchida2011}%
  \BibitemOpen
  \bibfield  {author} {\bibinfo {author} {\bibfnamefont {N.}~\bibnamefont
  {Uchida}}\ and\ \bibinfo {author} {\bibfnamefont {R.}~\bibnamefont
  {Golestanian}},\ }\bibfield  {title} {\bibinfo {title} {{Generic conditions
  for hydrodynamic synchronization}},\ }\href
  {https://doi.org/10.1103/PhysRevLett.106.058104} {\bibfield  {journal}
  {\bibinfo  {journal} {Phys. Rev. Lett.}\ }\textbf {\bibinfo {volume} {106}},\
  \bibinfo {pages} {058104} (\bibinfo {year} {2011})}\BibitemShut {NoStop}%
\bibitem [{\citenamefont {Uchida}(2011)}]{Uchida2011a}%
  \BibitemOpen
  \bibfield  {author} {\bibinfo {author} {\bibfnamefont {N.}~\bibnamefont
  {Uchida}},\ }\bibfield  {title} {\bibinfo {title} {{Many-Body Theory of
  Synchronization by Long-Range Interactions}},\ }\href
  {https://doi.org/10.1103/PhysRevLett.106.064101} {\bibfield  {journal}
  {\bibinfo  {journal} {Phys. Rev. Lett.}\ }\textbf {\bibinfo {volume} {106}},\
  \bibinfo {pages} {064101} (\bibinfo {year} {2011})}\BibitemShut {NoStop}%
\bibitem [{\citenamefont {Uchida}\ and\ \citenamefont
  {Golestanian}(2012)}]{Uchida2012}%
  \BibitemOpen
  \bibfield  {author} {\bibinfo {author} {\bibfnamefont {N.}~\bibnamefont
  {Uchida}}\ and\ \bibinfo {author} {\bibfnamefont {R.}~\bibnamefont
  {Golestanian}},\ }\bibfield  {title} {\bibinfo {title} {{Hydrodynamic
  synchronization between objects with cyclic rigid trajectories.}},\ }\href
  {https://doi.org/10.1140/epje/i2012-12135-5} {\bibfield  {journal} {\bibinfo
  {journal} {Eur. Phys. J. E}\ }\textbf {\bibinfo {volume} {35}},\ \bibinfo
  {pages} {9813} (\bibinfo {year} {2012})}\BibitemShut {NoStop}%
\bibitem [{\citenamefont {Izumida}\ \emph {et~al.}(2016)\citenamefont
  {Izumida}, \citenamefont {Kori},\ and\ \citenamefont {Seifert}}]{B2015}%
  \BibitemOpen
  \bibfield  {author} {\bibinfo {author} {\bibfnamefont {Y.}~\bibnamefont
  {Izumida}}, \bibinfo {author} {\bibfnamefont {H.}~\bibnamefont {Kori}},\ and\
  \bibinfo {author} {\bibfnamefont {U.}~\bibnamefont {Seifert}},\ }\bibfield
  {title} {\bibinfo {title} {{Energetics of synchronization in coupled
  oscillators rotating on circular trajectories}},\ }\href@noop {} {\bibfield
  {journal} {\bibinfo  {journal} {Phys. Rev. E}\ }\textbf {\bibinfo {volume}
  {94}},\ \bibinfo {pages} {052221} (\bibinfo {year} {2016})}\BibitemShut
  {NoStop}%
\bibitem [{\citenamefont {Quaranta}\ \emph {et~al.}(2015)\citenamefont
  {Quaranta}, \citenamefont {Aubin-Tam},\ and\ \citenamefont
  {Tam}}]{quaranta2015hydrodynamics}%
  \BibitemOpen
  \bibfield  {author} {\bibinfo {author} {\bibfnamefont {G.}~\bibnamefont
  {Quaranta}}, \bibinfo {author} {\bibfnamefont {M.-E.}\ \bibnamefont
  {Aubin-Tam}},\ and\ \bibinfo {author} {\bibfnamefont {D.}~\bibnamefont
  {Tam}},\ }\bibfield  {title} {\bibinfo {title} {Hydrodynamics versus
  intracellular coupling in the synchronization of eukaryotic flagella},\
  }\href@noop {} {\bibfield  {journal} {\bibinfo  {journal} {Phys. Rev. Lett.}\
  }\textbf {\bibinfo {volume} {115}},\ \bibinfo {pages} {238101} (\bibinfo
  {year} {2015})}\BibitemShut {NoStop}%
\bibitem [{\citenamefont {Wan}\ and\ \citenamefont
  {Goldstein}(2016)}]{wan2016coordinated}%
  \BibitemOpen
  \bibfield  {author} {\bibinfo {author} {\bibfnamefont {K.~Y.}\ \bibnamefont
  {Wan}}\ and\ \bibinfo {author} {\bibfnamefont {R.~E.}\ \bibnamefont
  {Goldstein}},\ }\bibfield  {title} {\bibinfo {title} {Coordinated beating of
  algal flagella is mediated by basal coupling},\ }\href@noop {} {\bibfield
  {journal} {\bibinfo  {journal} {Proc. Natl. Acad. Sci. U.S.A.}\ }\textbf
  {\bibinfo {volume} {113}},\ \bibinfo {pages} {E2784} (\bibinfo {year}
  {2016})}\BibitemShut {NoStop}%
\bibitem [{\citenamefont {Nitsan}\ \emph {et~al.}(2016)\citenamefont {Nitsan},
  \citenamefont {Drori}, \citenamefont {Lewis}, \citenamefont {Cohen},\ and\
  \citenamefont {Tzlil}}]{nitsan2016mechanical}%
  \BibitemOpen
  \bibfield  {author} {\bibinfo {author} {\bibfnamefont {I.}~\bibnamefont
  {Nitsan}}, \bibinfo {author} {\bibfnamefont {S.}~\bibnamefont {Drori}},
  \bibinfo {author} {\bibfnamefont {Y.~E.}\ \bibnamefont {Lewis}}, \bibinfo
  {author} {\bibfnamefont {S.}~\bibnamefont {Cohen}},\ and\ \bibinfo {author}
  {\bibfnamefont {S.}~\bibnamefont {Tzlil}},\ }\bibfield  {title} {\bibinfo
  {title} {Mechanical communication in cardiac cell synchronized beating},\
  }\href@noop {} {\bibfield  {journal} {\bibinfo  {journal} {Nat. Phys.}\
  }\textbf {\bibinfo {volume} {12}},\ \bibinfo {pages} {472} (\bibinfo {year}
  {2016})}\BibitemShut {NoStop}%
\bibitem [{\citenamefont {Cohen}\ and\ \citenamefont
  {Safran}(2016)}]{cohen2016elastic}%
  \BibitemOpen
  \bibfield  {author} {\bibinfo {author} {\bibfnamefont {O.}~\bibnamefont
  {Cohen}}\ and\ \bibinfo {author} {\bibfnamefont {S.~A.}\ \bibnamefont
  {Safran}},\ }\bibfield  {title} {\bibinfo {title} {Elastic interactions
  synchronize beating in cardiomyocytes},\ }\href@noop {} {\bibfield  {journal}
  {\bibinfo  {journal} {Soft Matter}\ }\textbf {\bibinfo {volume} {12}},\
  \bibinfo {pages} {6088} (\bibinfo {year} {2016})}\BibitemShut {NoStop}%
\bibitem [{\citenamefont {Magnasco}(1994)}]{Magnasco1994}%
  \BibitemOpen
  \bibfield  {author} {\bibinfo {author} {\bibfnamefont {M.~O.}\ \bibnamefont
  {Magnasco}},\ }\bibfield  {title} {\bibinfo {title} {{Molecular combustion
  motors}},\ }\href {https://doi.org/10.1103/PhysRevLett.72.2656} {\bibfield
  {journal} {\bibinfo  {journal} {Phys. Rev. Lett.}\ }\textbf {\bibinfo
  {volume} {72}},\ \bibinfo {pages} {2656} (\bibinfo {year}
  {1994})}\BibitemShut {NoStop}%
\bibitem [{\citenamefont {J{\"{u}}licher}\ \emph {et~al.}(1997)\citenamefont
  {J{\"{u}}licher}, \citenamefont {Ajdari},\ and\ \citenamefont
  {Prost}}]{Julicher1997}%
  \BibitemOpen
  \bibfield  {author} {\bibinfo {author} {\bibfnamefont {F.}~\bibnamefont
  {J{\"{u}}licher}}, \bibinfo {author} {\bibfnamefont {A.}~\bibnamefont
  {Ajdari}},\ and\ \bibinfo {author} {\bibfnamefont {J.}~\bibnamefont
  {Prost}},\ }\bibfield  {title} {\bibinfo {title} {{Modeling molecular
  motors}},\ }\href {https://doi.org/10.1103/RevModPhys.69.1269} {\bibfield
  {journal} {\bibinfo  {journal} {Rev. Mod. Phys.}\ }\textbf {\bibinfo {volume}
  {69}},\ \bibinfo {pages} {1269} (\bibinfo {year} {1997})}\BibitemShut
  {NoStop}%
\bibitem [{\citenamefont {Prost}\ \emph {et~al.}(1994)\citenamefont {Prost},
  \citenamefont {Chauwin}, \citenamefont {Peliti},\ and\ \citenamefont
  {Ajdari}}]{Prost1994}%
  \BibitemOpen
  \bibfield  {author} {\bibinfo {author} {\bibfnamefont {J.}~\bibnamefont
  {Prost}}, \bibinfo {author} {\bibfnamefont {J.~F.}\ \bibnamefont {Chauwin}},
  \bibinfo {author} {\bibfnamefont {L.}~\bibnamefont {Peliti}},\ and\ \bibinfo
  {author} {\bibfnamefont {A.}~\bibnamefont {Ajdari}},\ }\bibfield  {title}
  {\bibinfo {title} {{Asymmetric pumping of particles}},\ }\href
  {https://doi.org/10.1103/PhysRevLett.72.2652} {\bibfield  {journal} {\bibinfo
   {journal} {Phys. Rev. Lett.}\ }\textbf {\bibinfo {volume} {72}},\ \bibinfo
  {pages} {2652} (\bibinfo {year} {1994})}\BibitemShut {NoStop}%
\bibitem [{\citenamefont {Golubeva}\ \emph {et~al.}(2012)\citenamefont
  {Golubeva}, \citenamefont {Imparato},\ and\ \citenamefont
  {Peliti}}]{Golubeva2012}%
  \BibitemOpen
  \bibfield  {author} {\bibinfo {author} {\bibfnamefont {N.}~\bibnamefont
  {Golubeva}}, \bibinfo {author} {\bibfnamefont {A.}~\bibnamefont {Imparato}},\
  and\ \bibinfo {author} {\bibfnamefont {L.}~\bibnamefont {Peliti}},\
  }\bibfield  {title} {\bibinfo {title} {{Efficiency of molecular machines with
  continuous phase space}},\ }\href
  {https://doi.org/10.1209/0295-5075/97/60005} {\bibfield  {journal} {\bibinfo
  {journal} {EPL}\ }\textbf {\bibinfo {volume} {97}},\ \bibinfo {pages} {60005}
  (\bibinfo {year} {2012})}\BibitemShut {NoStop}%
\bibitem [{\citenamefont {Malgaretti}\ \emph {et~al.}(2012)\citenamefont
  {Malgaretti}, \citenamefont {Pagonabarraga},\ and\ \citenamefont
  {Frenkel}}]{Malgaretti2012}%
  \BibitemOpen
  \bibfield  {author} {\bibinfo {author} {\bibfnamefont {P.}~\bibnamefont
  {Malgaretti}}, \bibinfo {author} {\bibfnamefont {I.}~\bibnamefont
  {Pagonabarraga}},\ and\ \bibinfo {author} {\bibfnamefont {D.}~\bibnamefont
  {Frenkel}},\ }\bibfield  {title} {\bibinfo {title} {{Running Faster Together:
  Huge Speed up of Thermal Ratchets due to Hydrodynamic Coupling}},\
  }\href@noop {} {\bibfield  {journal} {\bibinfo  {journal} {Phys. Rev. Lett.}\
  }\textbf {\bibinfo {volume} {109}},\ \bibinfo {pages} {168101} (\bibinfo
  {year} {2012})}\BibitemShut {NoStop}%
\bibitem [{\citenamefont {Glowacki}\ \emph {et~al.}(2012)\citenamefont
  {Glowacki}, \citenamefont {Harvey},\ and\ \citenamefont
  {Mulholland}}]{glowacki2012taking}%
  \BibitemOpen
  \bibfield  {author} {\bibinfo {author} {\bibfnamefont {D.~R.}\ \bibnamefont
  {Glowacki}}, \bibinfo {author} {\bibfnamefont {J.~N.}\ \bibnamefont
  {Harvey}},\ and\ \bibinfo {author} {\bibfnamefont {A.~J.}\ \bibnamefont
  {Mulholland}},\ }\bibfield  {title} {\bibinfo {title} {Taking ockham's razor
  to enzyme dynamics and catalysis},\ }\href@noop {} {\bibfield  {journal}
  {\bibinfo  {journal} {Nat. Chem.}\ }\textbf {\bibinfo {volume} {4}},\
  \bibinfo {pages} {169} (\bibinfo {year} {2012})}\BibitemShut {NoStop}%
\bibitem [{\citenamefont {Callender}\ and\ \citenamefont
  {Dyer}(2015)}]{callender2015dynamical}%
  \BibitemOpen
  \bibfield  {author} {\bibinfo {author} {\bibfnamefont {R.}~\bibnamefont
  {Callender}}\ and\ \bibinfo {author} {\bibfnamefont {R.~B.}\ \bibnamefont
  {Dyer}},\ }\bibfield  {title} {\bibinfo {title} {The dynamical nature of
  enzymatic catalysis},\ }\href@noop {} {\bibfield  {journal} {\bibinfo
  {journal} {Acc. Chem. Res.}\ }\textbf {\bibinfo {volume} {48}},\ \bibinfo
  {pages} {407} (\bibinfo {year} {2015})}\BibitemShut {NoStop}%
\bibitem [{\citenamefont {G{\"{u}}nther}\ \emph {et~al.}(2018)\citenamefont
  {G{\"{u}}nther}, \citenamefont {B{\"{o}}rsch},\ and\ \citenamefont
  {Fischer}}]{Gunther2018a}%
  \BibitemOpen
  \bibfield  {author} {\bibinfo {author} {\bibfnamefont {J.~P.}\ \bibnamefont
  {G{\"{u}}nther}}, \bibinfo {author} {\bibfnamefont {M.}~\bibnamefont
  {B{\"{o}}rsch}},\ and\ \bibinfo {author} {\bibfnamefont {P.}~\bibnamefont
  {Fischer}},\ }\bibfield  {title} {\bibinfo {title} {{Diffusion Measurements
  of Swimming Enzymes with Fluorescence Correlation Spectroscopy}},\ }\href
  {https://doi.org/10.1021/acs.accounts.8b00276} {\bibfield  {journal}
  {\bibinfo  {journal} {Acc. Chem. Res.}\ }\textbf {\bibinfo {volume} {51}},\
  \bibinfo {pages} {1911} (\bibinfo {year} {2018})}\BibitemShut {NoStop}%
\bibitem [{\citenamefont {Xu}\ \emph {et~al.}(2019)\citenamefont {Xu},
  \citenamefont {Ross}, \citenamefont {Valdez},\ and\ \citenamefont
  {Sen}}]{Ross2019}%
  \BibitemOpen
  \bibfield  {author} {\bibinfo {author} {\bibfnamefont {M.}~\bibnamefont
  {Xu}}, \bibinfo {author} {\bibfnamefont {J.~L.}\ \bibnamefont {Ross}},
  \bibinfo {author} {\bibfnamefont {L.}~\bibnamefont {Valdez}},\ and\ \bibinfo
  {author} {\bibfnamefont {A.}~\bibnamefont {Sen}},\ }\bibfield  {title}
  {\bibinfo {title} {Direct single molecule imaging of enhanced enzyme
  diffusion},\ }\href {https://doi.org/10.1103/PhysRevLett.123.128101}
  {\bibfield  {journal} {\bibinfo  {journal} {Phys. Rev. Lett.}\ }\textbf
  {\bibinfo {volume} {123}},\ \bibinfo {pages} {128101} (\bibinfo {year}
  {2019})}\BibitemShut {NoStop}%
\bibitem [{\citenamefont {Golestanian}(2010)}]{RG2010}%
  \BibitemOpen
  \bibfield  {author} {\bibinfo {author} {\bibfnamefont {R.}~\bibnamefont
  {Golestanian}},\ }\bibfield  {title} {\bibinfo {title} {Synthetic
  mechanochemical molecular swimmer},\ }\href
  {https://doi.org/10.1103/PhysRevLett.105.018103} {\bibfield  {journal}
  {\bibinfo  {journal} {Phys. Rev. Lett.}\ }\textbf {\bibinfo {volume} {105}},\
  \bibinfo {pages} {018103} (\bibinfo {year} {2010})}\BibitemShut {NoStop}%
\bibitem [{\citenamefont {Golestanian}(2015)}]{Golestanian2015}%
  \BibitemOpen
  \bibfield  {author} {\bibinfo {author} {\bibfnamefont {R.}~\bibnamefont
  {Golestanian}},\ }\bibfield  {title} {\bibinfo {title} {{Enhanced diffusion
  of enzymes that catalyze exothermic reactions}},\ }\href
  {https://doi.org/10.1103/PhysRevLett.115.108102} {\bibfield  {journal}
  {\bibinfo  {journal} {Phys. Rev. Lett.}\ }\textbf {\bibinfo {volume} {115}},\
  \bibinfo {pages} {108102} (\bibinfo {year} {2015})}\BibitemShut {NoStop}%
\bibitem [{\citenamefont {Mikhailov}\ and\ \citenamefont
  {Kapral}(2015)}]{Mikhailov2015}%
  \BibitemOpen
  \bibfield  {author} {\bibinfo {author} {\bibfnamefont {A.~S.}\ \bibnamefont
  {Mikhailov}}\ and\ \bibinfo {author} {\bibfnamefont {R.}~\bibnamefont
  {Kapral}},\ }\bibfield  {title} {\bibinfo {title} {{Hydrodynamic collective
  effects of active protein machines in solution and lipid bilayers}},\ }\href
  {https://doi.org/10.1073/pnas.1506825112} {\bibfield  {journal} {\bibinfo
  {journal} {Proc. Natl. Acad. Sci. U.S.A.}\ }\textbf {\bibinfo {volume}
  {112}},\ \bibinfo {pages} {E3639} (\bibinfo {year} {2015})}\BibitemShut
  {NoStop}%
\bibitem [{\citenamefont {Bai}\ and\ \citenamefont
  {Wolynes}(2015)}]{bai2015hydrodynamics}%
  \BibitemOpen
  \bibfield  {author} {\bibinfo {author} {\bibfnamefont {X.}~\bibnamefont
  {Bai}}\ and\ \bibinfo {author} {\bibfnamefont {P.~G.}\ \bibnamefont
  {Wolynes}},\ }\bibfield  {title} {\bibinfo {title} {On the hydrodynamics of
  swimming enzymes},\ }\href@noop {} {\bibfield  {journal} {\bibinfo  {journal}
  {J. Chem. Phys.}\ }\textbf {\bibinfo {volume} {143}},\ \bibinfo {pages}
  {10B616\_1} (\bibinfo {year} {2015})}\BibitemShut {NoStop}%
\bibitem [{\citenamefont {Illien}\ \emph {et~al.}(2017)\citenamefont {Illien},
  \citenamefont {Adeleke-Larodo},\ and\ \citenamefont
  {Golestanian}}]{Illien2017b}%
  \BibitemOpen
  \bibfield  {author} {\bibinfo {author} {\bibfnamefont {P.}~\bibnamefont
  {Illien}}, \bibinfo {author} {\bibfnamefont {T.}~\bibnamefont
  {Adeleke-Larodo}},\ and\ \bibinfo {author} {\bibfnamefont {R.}~\bibnamefont
  {Golestanian}},\ }\bibfield  {title} {\bibinfo {title} {{Diffusion of an
  enzyme : The role of fluctuation- induced hydrodynamic coupling}},\ }\href
  {https://doi.org/10.1209/0295-5075/119/40002} {\bibfield  {journal} {\bibinfo
   {journal} {EPL}\ }\textbf {\bibinfo {volume} {119}},\ \bibinfo {pages}
  {40002} (\bibinfo {year} {2017})}\BibitemShut {NoStop}%
\bibitem [{\citenamefont {Agudo-Canalejo}\ \emph {et~al.}(2018)\citenamefont
  {Agudo-Canalejo}, \citenamefont {Adeleke-Larodo}, \citenamefont {Illien},\
  and\ \citenamefont {Golestanian}}]{Agudo-Canalejo2018a}%
  \BibitemOpen
  \bibfield  {author} {\bibinfo {author} {\bibfnamefont {J.}~\bibnamefont
  {Agudo-Canalejo}}, \bibinfo {author} {\bibfnamefont {T.}~\bibnamefont
  {Adeleke-Larodo}}, \bibinfo {author} {\bibfnamefont {P.}~\bibnamefont
  {Illien}},\ and\ \bibinfo {author} {\bibfnamefont {R.}~\bibnamefont
  {Golestanian}},\ }\bibfield  {title} {\bibinfo {title} {{Enhanced Diffusion
  and Chemotaxis at the Nanoscale}},\ }\href
  {https://doi.org/10.1021/acs.accounts.8b00280} {\bibfield  {journal}
  {\bibinfo  {journal} {Acc. Chem. Res.}\ }\textbf {\bibinfo {volume} {51}},\
  \bibinfo {pages} {2365} (\bibinfo {year} {2018})}\BibitemShut {NoStop}%
\bibitem [{\citenamefont {Hosaka}\ \emph {et~al.}(2020)\citenamefont {Hosaka},
  \citenamefont {Komura},\ and\ \citenamefont
  {Mikhailov}}]{hosaka2020mechanochemical}%
  \BibitemOpen
  \bibfield  {author} {\bibinfo {author} {\bibfnamefont {Y.}~\bibnamefont
  {Hosaka}}, \bibinfo {author} {\bibfnamefont {S.}~\bibnamefont {Komura}},\
  and\ \bibinfo {author} {\bibfnamefont {A.~S.}\ \bibnamefont {Mikhailov}},\
  }\bibfield  {title} {\bibinfo {title} {Mechanochemical enzymes and protein
  machines as hydrodynamic force dipoles: the active dimer model},\ }\href@noop
  {} {\bibfield  {journal} {\bibinfo  {journal} {Soft Matter}\ }\textbf
  {\bibinfo {volume} {16}},\ \bibinfo {pages} {10734} (\bibinfo {year}
  {2020})}\BibitemShut {NoStop}%
\bibitem [{\citenamefont {Koyano}\ \emph {et~al.}(2020)\citenamefont {Koyano},
  \citenamefont {Kitahata},\ and\ \citenamefont
  {Mikhailov}}]{koyano2020diffusion}%
  \BibitemOpen
  \bibfield  {author} {\bibinfo {author} {\bibfnamefont {Y.}~\bibnamefont
  {Koyano}}, \bibinfo {author} {\bibfnamefont {H.}~\bibnamefont {Kitahata}},\
  and\ \bibinfo {author} {\bibfnamefont {A.~S.}\ \bibnamefont {Mikhailov}},\
  }\bibfield  {title} {\bibinfo {title} {Diffusion in crowded colloids of
  particles cyclically changing their shapes},\ }\href@noop {} {\bibfield
  {journal} {\bibinfo  {journal} {EPL}\ }\textbf {\bibinfo {volume} {128}},\
  \bibinfo {pages} {40003} (\bibinfo {year} {2020})}\BibitemShut {NoStop}%
\bibitem [{\citenamefont {Sweetlove}\ and\ \citenamefont
  {Fernie}(2018)}]{sweetlove2018role}%
  \BibitemOpen
  \bibfield  {author} {\bibinfo {author} {\bibfnamefont {L.~J.}\ \bibnamefont
  {Sweetlove}}\ and\ \bibinfo {author} {\bibfnamefont {A.~R.}\ \bibnamefont
  {Fernie}},\ }\bibfield  {title} {\bibinfo {title} {The role of dynamic enzyme
  assemblies and substrate channelling in metabolic regulation},\ }\href@noop
  {} {\bibfield  {journal} {\bibinfo  {journal} {Nat. Commun.}\ }\textbf
  {\bibinfo {volume} {9}},\ \bibinfo {pages} {1} (\bibinfo {year}
  {2018})}\BibitemShut {NoStop}%
\bibitem [{\citenamefont {Kerfeld}\ \emph {et~al.}(2018)\citenamefont
  {Kerfeld}, \citenamefont {Aussignargues}, \citenamefont {Zarzycki},
  \citenamefont {Cai},\ and\ \citenamefont {Sutter}}]{kerfeld2018bacterial}%
  \BibitemOpen
  \bibfield  {author} {\bibinfo {author} {\bibfnamefont {C.~A.}\ \bibnamefont
  {Kerfeld}}, \bibinfo {author} {\bibfnamefont {C.}~\bibnamefont
  {Aussignargues}}, \bibinfo {author} {\bibfnamefont {J.}~\bibnamefont
  {Zarzycki}}, \bibinfo {author} {\bibfnamefont {F.}~\bibnamefont {Cai}},\ and\
  \bibinfo {author} {\bibfnamefont {M.}~\bibnamefont {Sutter}},\ }\bibfield
  {title} {\bibinfo {title} {Bacterial microcompartments},\ }\href@noop {}
  {\bibfield  {journal} {\bibinfo  {journal} {Nat. Rev. Microbiol}\ }\textbf
  {\bibinfo {volume} {16}},\ \bibinfo {pages} {277} (\bibinfo {year}
  {2018})}\BibitemShut {NoStop}%
\bibitem [{\citenamefont {Liu}(2016)}]{liu2016cytoophidium}%
  \BibitemOpen
  \bibfield  {author} {\bibinfo {author} {\bibfnamefont {J.-L.}\ \bibnamefont
  {Liu}},\ }\bibfield  {title} {\bibinfo {title} {The cytoophidium and its
  kind: filamentation and compartmentation of metabolic enzymes},\ }\href@noop
  {} {\bibfield  {journal} {\bibinfo  {journal} {Annu. Rev. Cell Dev. Biol.}\
  }\textbf {\bibinfo {volume} {32}},\ \bibinfo {pages} {349} (\bibinfo {year}
  {2016})}\BibitemShut {NoStop}%
\bibitem [{\citenamefont {Selwood}\ and\ \citenamefont
  {Jaffe}(2012)}]{selwood2012dynamic}%
  \BibitemOpen
  \bibfield  {author} {\bibinfo {author} {\bibfnamefont {T.}~\bibnamefont
  {Selwood}}\ and\ \bibinfo {author} {\bibfnamefont {E.~K.}\ \bibnamefont
  {Jaffe}},\ }\bibfield  {title} {\bibinfo {title} {Dynamic dissociating
  homo-oligomers and the control of protein function},\ }\href@noop {}
  {\bibfield  {journal} {\bibinfo  {journal} {Arch. Biochem. Biophys.}\
  }\textbf {\bibinfo {volume} {519}},\ \bibinfo {pages} {131} (\bibinfo {year}
  {2012})}\BibitemShut {NoStop}%
\bibitem [{\citenamefont {Gomperts}(1996)}]{gomperts1996clustering}%
  \BibitemOpen
  \bibfield  {author} {\bibinfo {author} {\bibfnamefont {S.~N.}\ \bibnamefont
  {Gomperts}},\ }\bibfield  {title} {\bibinfo {title} {Clustering membrane
  proteins: It's all coming together with the psd-95/sap90 protein family},\
  }\href@noop {} {\bibfield  {journal} {\bibinfo  {journal} {Cell}\ }\textbf
  {\bibinfo {volume} {84}},\ \bibinfo {pages} {659} (\bibinfo {year}
  {1996})}\BibitemShut {NoStop}%
\bibitem [{\citenamefont {Shuai}\ and\ \citenamefont
  {Jung}(2003)}]{shuai2003optimal}%
  \BibitemOpen
  \bibfield  {author} {\bibinfo {author} {\bibfnamefont {J.}~\bibnamefont
  {Shuai}}\ and\ \bibinfo {author} {\bibfnamefont {P.}~\bibnamefont {Jung}},\
  }\bibfield  {title} {\bibinfo {title} {Optimal ion channel clustering for
  intracellular calcium signaling},\ }\href@noop {} {\bibfield  {journal}
  {\bibinfo  {journal} {Proc. Natl. Acad. Sci. U.S.A}\ }\textbf {\bibinfo
  {volume} {100}},\ \bibinfo {pages} {506} (\bibinfo {year}
  {2003})}\BibitemShut {NoStop}%
\bibitem [{sup()}]{suppmat}%
  \BibitemOpen
  \href@noop {} {}\bibinfo {note} {See Supplemental Material at [URL will be
  inserted by publisher] for details on the calculation of $h$ for bound
  enzymes, the elimination of fast degrees of freedom, the construction of the
  stochastic dynamics, and Supplemental Figures S1--S3.}\BibitemShut {Stop}%
\bibitem [{non()}]{nonzeromu}%
  \BibitemOpen
  \href@noop {} {}\bibinfo {note} {We believe this to be a reasonable
  assumption. If the conformational changes were the bottleneck for the
  dynamics, even the catalytic activity of a single enzyme would be strongly
  affected by its conformational dynamics, which does not seem to be the case
  in practice, see e.g.~Ref.~\cite{kamerlin2010dawn}. Moreover, in Fig.~S2 in
  the Supplemental Material \cite{suppmat} we show that the same type of
  bifurcation occurs when this assumption is relaxed.}\BibitemShut {Stop}%
\bibitem [{con()}]{constantcoupling}%
  \BibitemOpen
  \href@noop {} {}\bibinfo {note} {As an example, in Fig.~S3 in the
  Supplemental Material \cite{suppmat} we show that the same type of
  bifurcation occurs in a simpler model in which the off-diagonal coefficients
  of the mobility matrix are constant.}\BibitemShut {Stop}%
\bibitem [{\citenamefont {Lindner}\ \emph {et~al.}(2004)\citenamefont
  {Lindner}, \citenamefont {Garc{\i}a-Ojalvo}, \citenamefont {Neiman},\ and\
  \citenamefont {Schimansky-Geier}}]{lindner2004effects}%
  \BibitemOpen
  \bibfield  {author} {\bibinfo {author} {\bibfnamefont {B.}~\bibnamefont
  {Lindner}}, \bibinfo {author} {\bibfnamefont {J.}~\bibnamefont
  {Garc{\i}a-Ojalvo}}, \bibinfo {author} {\bibfnamefont {A.}~\bibnamefont
  {Neiman}},\ and\ \bibinfo {author} {\bibfnamefont {L.}~\bibnamefont
  {Schimansky-Geier}},\ }\bibfield  {title} {\bibinfo {title} {Effects of noise
  in excitable systems},\ }\href@noop {} {\bibfield  {journal} {\bibinfo
  {journal} {Phys. Rep.}\ }\textbf {\bibinfo {volume} {392}},\ \bibinfo {pages}
  {321} (\bibinfo {year} {2004})}\BibitemShut {NoStop}%
\bibitem [{\citenamefont {Neiman}\ \emph {et~al.}(1999)\citenamefont {Neiman},
  \citenamefont {Schimansky-Geier}, \citenamefont {Cornell-Bell},\ and\
  \citenamefont {Moss}}]{neiman1999noise}%
  \BibitemOpen
  \bibfield  {author} {\bibinfo {author} {\bibfnamefont {A.}~\bibnamefont
  {Neiman}}, \bibinfo {author} {\bibfnamefont {L.}~\bibnamefont
  {Schimansky-Geier}}, \bibinfo {author} {\bibfnamefont {A.}~\bibnamefont
  {Cornell-Bell}},\ and\ \bibinfo {author} {\bibfnamefont {F.}~\bibnamefont
  {Moss}},\ }\bibfield  {title} {\bibinfo {title} {Noise-enhanced phase
  synchronization in excitable media},\ }\href@noop {} {\bibfield  {journal}
  {\bibinfo  {journal} {Phys. Rev. Lett.}\ }\textbf {\bibinfo {volume} {83}},\
  \bibinfo {pages} {4896} (\bibinfo {year} {1999})}\BibitemShut {NoStop}%
\bibitem [{\citenamefont {Hu}\ and\ \citenamefont {Zhou}(2000)}]{hu2000phase}%
  \BibitemOpen
  \bibfield  {author} {\bibinfo {author} {\bibfnamefont {B.}~\bibnamefont
  {Hu}}\ and\ \bibinfo {author} {\bibfnamefont {C.}~\bibnamefont {Zhou}},\
  }\bibfield  {title} {\bibinfo {title} {Phase synchronization in coupled
  nonidentical excitable systems and array-enhanced coherence resonance},\
  }\href@noop {} {\bibfield  {journal} {\bibinfo  {journal} {Phys. Rev. E}\
  }\textbf {\bibinfo {volume} {61}},\ \bibinfo {pages} {R1001} (\bibinfo {year}
  {2000})}\BibitemShut {NoStop}%
\bibitem [{\citenamefont {Sosnovtseva}\ \emph {et~al.}(2001)\citenamefont
  {Sosnovtseva}, \citenamefont {Fomin}, \citenamefont {Postnov},\ and\
  \citenamefont {Anishchenko}}]{sosnovtseva2001clustering}%
  \BibitemOpen
  \bibfield  {author} {\bibinfo {author} {\bibfnamefont {O.}~\bibnamefont
  {Sosnovtseva}}, \bibinfo {author} {\bibfnamefont {A.}~\bibnamefont {Fomin}},
  \bibinfo {author} {\bibfnamefont {D.}~\bibnamefont {Postnov}},\ and\ \bibinfo
  {author} {\bibfnamefont {V.}~\bibnamefont {Anishchenko}},\ }\bibfield
  {title} {\bibinfo {title} {Clustering of noise-induced oscillations},\
  }\href@noop {} {\bibfield  {journal} {\bibinfo  {journal} {Phys. Rev. E}\
  }\textbf {\bibinfo {volume} {64}},\ \bibinfo {pages} {026204} (\bibinfo
  {year} {2001})}\BibitemShut {NoStop}%
\bibitem [{\citenamefont {Woillez}\ \emph {et~al.}(2019)\citenamefont
  {Woillez}, \citenamefont {Zhao}, \citenamefont {Kafri}, \citenamefont
  {Lecomte},\ and\ \citenamefont {Tailleur}}]{PhysRevLett.122.258001}%
  \BibitemOpen
  \bibfield  {author} {\bibinfo {author} {\bibfnamefont {E.}~\bibnamefont
  {Woillez}}, \bibinfo {author} {\bibfnamefont {Y.}~\bibnamefont {Zhao}},
  \bibinfo {author} {\bibfnamefont {Y.}~\bibnamefont {Kafri}}, \bibinfo
  {author} {\bibfnamefont {V.}~\bibnamefont {Lecomte}},\ and\ \bibinfo {author}
  {\bibfnamefont {J.}~\bibnamefont {Tailleur}},\ }\bibfield  {title} {\bibinfo
  {title} {Activated escape of a self-propelled particle from a metastable
  state},\ }\href {https://doi.org/10.1103/PhysRevLett.122.258001} {\bibfield
  {journal} {\bibinfo  {journal} {Phys. Rev. Lett.}\ }\textbf {\bibinfo
  {volume} {122}},\ \bibinfo {pages} {258001} (\bibinfo {year}
  {2019})}\BibitemShut {NoStop}%
\bibitem [{\citenamefont {Woillez}\ \emph
  {et~al.}(2020{\natexlab{a}})\citenamefont {Woillez}, \citenamefont {Kafri},\
  and\ \citenamefont {Gov}}]{PhysRevLett.124.118002}%
  \BibitemOpen
  \bibfield  {author} {\bibinfo {author} {\bibfnamefont {E.}~\bibnamefont
  {Woillez}}, \bibinfo {author} {\bibfnamefont {Y.}~\bibnamefont {Kafri}},\
  and\ \bibinfo {author} {\bibfnamefont {N.~S.}\ \bibnamefont {Gov}},\
  }\bibfield  {title} {\bibinfo {title} {Active trap model},\ }\href
  {https://doi.org/10.1103/PhysRevLett.124.118002} {\bibfield  {journal}
  {\bibinfo  {journal} {Phys. Rev. Lett.}\ }\textbf {\bibinfo {volume} {124}},\
  \bibinfo {pages} {118002} (\bibinfo {year} {2020}{\natexlab{a}})}\BibitemShut
  {NoStop}%
\bibitem [{\citenamefont {Woillez}\ \emph
  {et~al.}(2020{\natexlab{b}})\citenamefont {Woillez}, \citenamefont {Kafri},\
  and\ \citenamefont {Lecomte}}]{woillez2020nonlocal}%
  \BibitemOpen
  \bibfield  {author} {\bibinfo {author} {\bibfnamefont {E.}~\bibnamefont
  {Woillez}}, \bibinfo {author} {\bibfnamefont {Y.}~\bibnamefont {Kafri}},\
  and\ \bibinfo {author} {\bibfnamefont {V.}~\bibnamefont {Lecomte}},\
  }\bibfield  {title} {\bibinfo {title} {Nonlocal stationary probability
  distributions and escape rates for an active ornstein--uhlenbeck particle},\
  }\href@noop {} {\bibfield  {journal} {\bibinfo  {journal} {J. Stat. Mech.:
  Theory Exp.}\ }\textbf {\bibinfo {volume} {2020}}\bibinfo  {number} { (6)},\
  \bibinfo {pages} {063204}}\BibitemShut {NoStop}%
\bibitem [{\citenamefont {Dueber}\ \emph {et~al.}(2009)\citenamefont {Dueber},
  \citenamefont {Wu}, \citenamefont {Malmirchegini}, \citenamefont {Moon},
  \citenamefont {Petzold}, \citenamefont {Ullal}, \citenamefont {Prather},\
  and\ \citenamefont {Keasling}}]{dueber2009synthetic}%
  \BibitemOpen
\bibfield  {number} {  }\bibfield  {author} {\bibinfo {author} {\bibfnamefont
  {J.~E.}\ \bibnamefont {Dueber}}, \bibinfo {author} {\bibfnamefont {G.~C.}\
  \bibnamefont {Wu}}, \bibinfo {author} {\bibfnamefont {G.~R.}\ \bibnamefont
  {Malmirchegini}}, \bibinfo {author} {\bibfnamefont {T.~S.}\ \bibnamefont
  {Moon}}, \bibinfo {author} {\bibfnamefont {C.~J.}\ \bibnamefont {Petzold}},
  \bibinfo {author} {\bibfnamefont {A.~V.}\ \bibnamefont {Ullal}}, \bibinfo
  {author} {\bibfnamefont {K.~L.}\ \bibnamefont {Prather}},\ and\ \bibinfo
  {author} {\bibfnamefont {J.~D.}\ \bibnamefont {Keasling}},\ }\bibfield
  {title} {\bibinfo {title} {Synthetic protein scaffolds provide modular
  control over metabolic flux},\ }\href@noop {} {\bibfield  {journal} {\bibinfo
   {journal} {Nat. Biotechnol.}\ }\textbf {\bibinfo {volume} {27}},\ \bibinfo
  {pages} {753} (\bibinfo {year} {2009})}\BibitemShut {NoStop}%
\bibitem [{\citenamefont {Kufer}\ \emph {et~al.}(2008)\citenamefont {Kufer},
  \citenamefont {Puchner}, \citenamefont {Gumpp}, \citenamefont {Liedl},\ and\
  \citenamefont {Gaub}}]{kufer2008single}%
  \BibitemOpen
  \bibfield  {author} {\bibinfo {author} {\bibfnamefont {S.}~\bibnamefont
  {Kufer}}, \bibinfo {author} {\bibfnamefont {E.}~\bibnamefont {Puchner}},
  \bibinfo {author} {\bibfnamefont {H.}~\bibnamefont {Gumpp}}, \bibinfo
  {author} {\bibfnamefont {T.}~\bibnamefont {Liedl}},\ and\ \bibinfo {author}
  {\bibfnamefont {H.}~\bibnamefont {Gaub}},\ }\bibfield  {title} {\bibinfo
  {title} {Single-molecule cut-and-paste surface assembly},\ }\href@noop {}
  {\bibfield  {journal} {\bibinfo  {journal} {Science}\ }\textbf {\bibinfo
  {volume} {319}},\ \bibinfo {pages} {594} (\bibinfo {year}
  {2008})}\BibitemShut {NoStop}%
\bibitem [{\citenamefont {Wilner}\ \emph {et~al.}(2009)\citenamefont {Wilner},
  \citenamefont {Weizmann}, \citenamefont {Gill}, \citenamefont
  {Lioubashevski}, \citenamefont {Freeman},\ and\ \citenamefont
  {Willner}}]{wilner2009enzyme}%
  \BibitemOpen
  \bibfield  {author} {\bibinfo {author} {\bibfnamefont {O.~I.}\ \bibnamefont
  {Wilner}}, \bibinfo {author} {\bibfnamefont {Y.}~\bibnamefont {Weizmann}},
  \bibinfo {author} {\bibfnamefont {R.}~\bibnamefont {Gill}}, \bibinfo {author}
  {\bibfnamefont {O.}~\bibnamefont {Lioubashevski}}, \bibinfo {author}
  {\bibfnamefont {R.}~\bibnamefont {Freeman}},\ and\ \bibinfo {author}
  {\bibfnamefont {I.}~\bibnamefont {Willner}},\ }\bibfield  {title} {\bibinfo
  {title} {Enzyme cascades activated on topologically programmed dna
  scaffolds},\ }\href@noop {} {\bibfield  {journal} {\bibinfo  {journal} {Nat.
  Nanotechnol.}\ }\textbf {\bibinfo {volume} {4}},\ \bibinfo {pages} {249}
  (\bibinfo {year} {2009})}\BibitemShut {NoStop}%
\bibitem [{\citenamefont {Min}\ \emph {et~al.}(2005)\citenamefont {Min},
  \citenamefont {English}, \citenamefont {Luo}, \citenamefont {Cherayil},
  \citenamefont {Kou},\ and\ \citenamefont {Xie}}]{min2005fluctuating}%
  \BibitemOpen
  \bibfield  {author} {\bibinfo {author} {\bibfnamefont {W.}~\bibnamefont
  {Min}}, \bibinfo {author} {\bibfnamefont {B.~P.}\ \bibnamefont {English}},
  \bibinfo {author} {\bibfnamefont {G.}~\bibnamefont {Luo}}, \bibinfo {author}
  {\bibfnamefont {B.~J.}\ \bibnamefont {Cherayil}}, \bibinfo {author}
  {\bibfnamefont {S.}~\bibnamefont {Kou}},\ and\ \bibinfo {author}
  {\bibfnamefont {X.~S.}\ \bibnamefont {Xie}},\ }\bibfield  {title} {\bibinfo
  {title} {Fluctuating enzymes: lessons from single-molecule studies},\
  }\href@noop {} {\bibfield  {journal} {\bibinfo  {journal} {Acc. Chem. Res.}\
  }\textbf {\bibinfo {volume} {38}},\ \bibinfo {pages} {923} (\bibinfo {year}
  {2005})}\BibitemShut {NoStop}%
\bibitem [{\citenamefont {Gumpp}\ \emph {et~al.}(2009)\citenamefont {Gumpp},
  \citenamefont {Puchner}, \citenamefont {Zimmermann}, \citenamefont {Gerland},
  \citenamefont {Gaub},\ and\ \citenamefont {Blank}}]{gumpp2009triggering}%
  \BibitemOpen
  \bibfield  {author} {\bibinfo {author} {\bibfnamefont {H.}~\bibnamefont
  {Gumpp}}, \bibinfo {author} {\bibfnamefont {E.~M.}\ \bibnamefont {Puchner}},
  \bibinfo {author} {\bibfnamefont {J.~L.}\ \bibnamefont {Zimmermann}},
  \bibinfo {author} {\bibfnamefont {U.}~\bibnamefont {Gerland}}, \bibinfo
  {author} {\bibfnamefont {H.~E.}\ \bibnamefont {Gaub}},\ and\ \bibinfo
  {author} {\bibfnamefont {K.}~\bibnamefont {Blank}},\ }\bibfield  {title}
  {\bibinfo {title} {Triggering enzymatic activity with force},\ }\href@noop {}
  {\bibfield  {journal} {\bibinfo  {journal} {Nano Lett.}\ }\textbf {\bibinfo
  {volume} {9}},\ \bibinfo {pages} {3290} (\bibinfo {year} {2009})}\BibitemShut
  {NoStop}%
\bibitem [{\citenamefont {Choi}\ \emph {et~al.}(2012)\citenamefont {Choi},
  \citenamefont {Moody}, \citenamefont {Sims}, \citenamefont {Hunt},
  \citenamefont {Corso}, \citenamefont {Perez}, \citenamefont {Weiss},\ and\
  \citenamefont {Collins}}]{choi2012single}%
  \BibitemOpen
  \bibfield  {author} {\bibinfo {author} {\bibfnamefont {Y.}~\bibnamefont
  {Choi}}, \bibinfo {author} {\bibfnamefont {I.~S.}\ \bibnamefont {Moody}},
  \bibinfo {author} {\bibfnamefont {P.~C.}\ \bibnamefont {Sims}}, \bibinfo
  {author} {\bibfnamefont {S.~R.}\ \bibnamefont {Hunt}}, \bibinfo {author}
  {\bibfnamefont {B.~L.}\ \bibnamefont {Corso}}, \bibinfo {author}
  {\bibfnamefont {I.}~\bibnamefont {Perez}}, \bibinfo {author} {\bibfnamefont
  {G.~A.}\ \bibnamefont {Weiss}},\ and\ \bibinfo {author} {\bibfnamefont
  {P.~G.}\ \bibnamefont {Collins}},\ }\bibfield  {title} {\bibinfo {title}
  {Single-molecule lysozyme dynamics monitored by an electronic circuit},\
  }\href@noop {} {\bibfield  {journal} {\bibinfo  {journal} {Science}\ }\textbf
  {\bibinfo {volume} {335}},\ \bibinfo {pages} {319} (\bibinfo {year}
  {2012})}\BibitemShut {NoStop}%
\bibitem [{\citenamefont {Pelz}\ \emph {et~al.}(2016)\citenamefont {Pelz},
  \citenamefont {{\v{Z}}old{\'a}k}, \citenamefont {Zeller}, \citenamefont
  {Zacharias},\ and\ \citenamefont {Rief}}]{pelz2016subnanometre}%
  \BibitemOpen
  \bibfield  {author} {\bibinfo {author} {\bibfnamefont {B.}~\bibnamefont
  {Pelz}}, \bibinfo {author} {\bibfnamefont {G.}~\bibnamefont
  {{\v{Z}}old{\'a}k}}, \bibinfo {author} {\bibfnamefont {F.}~\bibnamefont
  {Zeller}}, \bibinfo {author} {\bibfnamefont {M.}~\bibnamefont {Zacharias}},\
  and\ \bibinfo {author} {\bibfnamefont {M.}~\bibnamefont {Rief}},\ }\bibfield
  {title} {\bibinfo {title} {Subnanometre enzyme mechanics probed by
  single-molecule force spectroscopy},\ }\href@noop {} {\bibfield  {journal}
  {\bibinfo  {journal} {Nat. Commun.}\ }\textbf {\bibinfo {volume} {7}},\
  \bibinfo {pages} {1} (\bibinfo {year} {2016})}\BibitemShut {NoStop}%
\bibitem [{\citenamefont {Kamerlin}\ and\ \citenamefont
  {Warshel}(2010)}]{kamerlin2010dawn}%
  \BibitemOpen
  \bibfield  {author} {\bibinfo {author} {\bibfnamefont {S.~C.}\ \bibnamefont
  {Kamerlin}}\ and\ \bibinfo {author} {\bibfnamefont {A.}~\bibnamefont
  {Warshel}},\ }\bibfield  {title} {\bibinfo {title} {At the dawn of the 21st
  century: Is dynamics the missing link for understanding enzyme catalysis?},\
  }\href@noop {} {\bibfield  {journal} {\bibinfo  {journal} {Proteins}\
  }\textbf {\bibinfo {volume} {78}},\ \bibinfo {pages} {1339} (\bibinfo {year}
  {2010})}\BibitemShut {NoStop}%
\end{thebibliography}%

\end{document}


\title{Synchronization and enhanced catalysis of mechanically coupled enzymes \\ {\it Supplemental Material}}

\author{Jaime Agudo-Canalejo}
\affiliation{Department of Living Matter Physics, Max Planck Institute for Dynamics and Self-Organization, D-37077 G\"ottingen, Germany}

\author{Tunrayo Adeleke-Larodo}
\affiliation{Rudolf Peierls Centre for Theoretical Physics, University of Oxford, Oxford OX1 3PU, United Kingdom}

\author{Pierre Illien}
\affiliation{Sorbonne Universit\'e, CNRS, Laboratoire Physicochimie des Electrolytes et Nanosyst\`emes Interfaciaux (PHENIX), UMR  8234, 4 place Jussieu, 75005 Paris, France}

\author{Ramin Golestanian}
\affiliation{Department of Living Matter Physics, Max Planck Institute for Dynamics and Self-Organization, D-37077 G\"ottingen, Germany}
\affiliation{Rudolf Peierls Centre for Theoretical Physics, University of Oxford, Oxford OX1 3PU, United Kingdom}

\date{\today}

\maketitle

\renewcommand{\theequation}{S\arabic{equation}}
\renewcommand{\thefigure}{S\arabic{figure}}

\section*{Coupling constant for a two-enzyme complex}

We consider two identical enzymes that are directly bound to each other, forming a complex, as in Fig.~1(a) of the main text. Assuming that the link between the two enzymes is perfectly rigid, the full complex can be divided into three compact subunits. From left to right in Fig.~1(a), we name them A1 (the part of enzyme 1 not bound to enzyme 2), B (the parts of enzymes 1 and 2 that are bound together), and A2 (the part of enzyme 2 not bound to enzyme 1). Each subunit will have positions given by $\rr_{A1}$, $\rr_B$, and $\rr_{A2}$, and hydrodynamic mobilities $\mu_{A1}$, $\mu_{B}$, and $\mu_{A2}$, respectively. Because the two enzymes are identical, we set $\mu_{A1}=\mu_{A2} \equiv \mu_A$.

We define the orientation of the complex as $\nn \equiv (\rr_B - \rr_{A1})/|\rr_B - \rr_{A1}| = (\rr_{A2} - \rr_{B})/|\rr_{A2} - \rr_{B}| = (\rr_{A2} - \rr_{A1})/|\rr_{A2} - \rr_{A1}|$. With this, the overdamped equations of motion for the positions of each subunit can be written as
\begin{eqnarray}
	\dot{\rr}_{A1} & = & - \mu_A f_1 \nn, \\
	\dot{\rr}_B & = & \mu_B (f_1 - f_2) \nn, \\
	\dot{\rr}_{A2} & = & \mu_A f_2 \nn.
\end{eqnarray}
From these, we can calculate
\begin{eqnarray}
	\dot{\rr}_B - \dot{\rr}_{A1} & = & [(\mu_A + \mu_B) f_1 - \mu_B f_2] \nn, \\
	\dot{\rr}_{A2} - \dot{\rr}_B & = & [(\mu_A + \mu_B) f_2 - \mu_B f_1] \nn.
\end{eqnarray}
Noting that $\rr_B - \rr_{A1} = L_1 \nn$ and $\rr_{A2} - \rr_B = L_2 \nn$, this finally gives
\begin{eqnarray}
	\dot{L}_1 & = & \mu ( f_1 + h f_2 ), \label{eq:L1} \\
	\dot{L}_2 & = & \mu ( f_2 + h f_1 ), \label{eq:L2}
\end{eqnarray}
where we have defined
\begin{eqnarray}
	\mu & \equiv & \mu_A+\mu_B, \\
	h & \equiv & - \frac{\mu_B}{\mu_A+\mu_B}. \label{eq:h}
\end{eqnarray}
Note that (\ref{eq:L1}-\ref{eq:L2}) have the same form as the force-velocity relations given in the main text. Moreover, $h$ is always negative and constrained to $0 > h > -1$, as mentioned in the main text: the limit $h \approx 0$ corresponds to $\mu_B \ll \mu_A$, while the limit $h \approx -1$ corresponds to $\mu_B \gg \mu_A$.

\section*{Projection of fast degrees of freedom onto slow manifold}

We start with the equations for the dynamics of the lengths $L_\alpha$ and phases $\phi_\alpha$ derived in the main text:
\begin{equation}
\label{eq:ldot_1}
\dot{L}_\alpha=\mu \big(f_\alpha + h f_\beta \big), 
\end{equation}
with $f_\alpha = - k [L_\alpha - L(\phi_\alpha)]$, and
\begin{equation}
\dot{\phi}_\alpha = -\mu_\phi \left\{ - k [L_\alpha - L(\phi_\alpha)] L'(\phi_\alpha) + V'(\phi_\alpha)  \right\}.
\label{eq:phi1}
\end{equation}

Equations (\ref{eq:ldot_1}) and (\ref{eq:phi1}) describe the coupled dynamics of length and phase. The dynamics can be simplified further if we assume that the enzyme is very rigid, so that the typical timescale of length relaxations $(\mu k)^{-1}$ is much shorter than that of changes in the phase velocity. The dynamics of the deviation of the length from its preferred value $\delta L_\alpha \equiv L_\alpha - L(\phi_\alpha)$ is governed by
\begin{equation}
L'(\phi_\alpha) \dot{\phi}_\alpha + \delta \dot{L}_\alpha =\dot{L}_\alpha= - \mu  k \big(\delta L_\alpha + h  \delta L_\beta \big).
\end{equation}
Assuming fast relaxation, we can set $\delta \dot{L}_\alpha = 0$ and solve for $\delta L_\alpha$ to obtain
\begin{equation}
 -\mu k \delta L_\alpha = \frac{L'(\phi_\alpha) \dot{\phi}_\alpha - h L'(\phi_\beta)\dot{\phi}_\beta}{1-h^2}. 
\label{eq:deviation}
\end{equation}
This shows that there is a difference between the elastic rest length of the enzyme and its dynamical rest length. The non-vanishing deformation leads to a finite force (-dipole) that is present during the enzyme activity.

Introducing Eq.~\eqref{eq:deviation} into Eq.~\eqref{eq:phi1}, and solving for $\dot{\phi}_\alpha$, we obtain
\begin{equation}
\dot{\phi}_\alpha = M_{\alpha \beta} [-\mu_\phi V'(\phi_\beta)],
\label{eq:fullphi}
\end{equation}
where Einstein summation convention for repeated indices is used and
\begin{eqnarray}
	M_{11} & = & \left[ 1+\frac{A_1^2}{1-h^2} - \left( \frac{h A_1 A_2}{1-h^2} \right)^2 \left(1+\frac{A_2^2}{1-h^2}\right)^{-1} \right]^{-1}, \label{eq:mob1} \\
	M_{22} & = & \left[ 1+\frac{A_2^2}{1-h^2} - \left( \frac{h A_1 A_2}{1-h^2} \right)^2 \left(1+\frac{A_1^2}{1-h^2}\right)^{-1} \right]^{-1}, \\
	M_{12} & = & M_{21} = \frac{h A_1 A_2}{1-h^2} \left[ \left( 1+\frac{A_1^2}{1-h^2} \right) \left( 1+\frac{A_2^2}{1-h^2} \right) - \left( \frac{h A_1 A_2}{1-h^2} \right)^2 \right]^{-1}, \label{eq:mob12}
\end{eqnarray}
with the definition $A_\alpha \equiv \sqrt{\frac{\mu_\phi}{\mu}} L'(\phi_\alpha)$. Importantly, the dynamics is independent of the stiffness $k$, as long as $k$ is large enough for the fast relaxation approximation to hold.

There are two relevant limits in which the expressions (\ref{eq:mob1}--\ref{eq:mob12}) simplify significantly. First, when the mobility of the phase is smaller than the mobility of the elongation, with $\frac{\mu_\phi}{\mu} L'(\phi_\alpha)^2 \ll 1$ and thus $A_\alpha \ll 1$, the phase mobility constitutes the bottleneck of the dynamics and  (\ref{eq:mob1}--\ref{eq:mob12}) become, to lowest order in $A_\alpha$, 
	\begin{eqnarray}
		M_{11} & = & M_{22} = 1, \label{eq:mob1b} \\
		M_{12} & = & M_{21} = \frac{h A_1 A_2}{1-h^2}. \label{eq:mob12b}
	\end{eqnarray}
This is the limit considered in the main text, see Eq.~(1) there, which is equivalent to (\ref{eq:fullphi}) with mobility matrix given by (\ref{eq:mob1b}-\ref{eq:mob12b}).

Second, in the limit of small coupling, without any assumptions on the relative values of the mobilities of phase and elongation,  (\ref{eq:mob1}--\ref{eq:mob12}) become, to lowest order in $h$,
\begin{eqnarray}
	M_{11} & = & \frac{1}{1+A_1^2}, \label{eq:mob1c} \\
	M_{22} & = & \frac{1}{1+A_2^2},  \\
	M_{12} & = & M_{21} = \frac{h A_1 A_2}{(1+A_1^2) (1+A_2^2)}.   \label{eq:mob12c}
\end{eqnarray}

\section*{Construction of the stochastic dynamics}

Since enzymes operate at the molecular scale, the appropriate description of their dynamics should include stochastic effects. To construct such a description, we would need to coarse-grain the stochastic dynamics of all the relevant degrees of freedom, which can in general be cumbersome. The systematic dimensional reduction of the dynamics in terms of the slow variables and the fact that in this limit $U(L_\alpha,\phi_\alpha) \simeq V(\phi_\alpha)$, however, provide a route to constructing the effective stochastic dynamics. 
To this end, starting from Eq.~(\ref{eq:fullphi}), we build the stochastic dynamics of the system in form of the corresponding Fokker-Planck equation 
\begin{equation}
\partial_t {\cal P} = \partial_{\alpha} \left[M_{\alpha \beta} \mu_\phi \Big( V'(\phi_\beta) {\cal P}+k_\mathrm{B}T  \partial_{\beta}{\cal P}\Big)\right],\label{eq:FP-def}
\end{equation}
for the probability distribution ${\cal P}(\phi_1,\phi_2;t)$, which ensures equilibration to the Boltzmann distribution ${\cal{P}}(\phi_1,\phi_2)\propto \exp\{-{[V(\phi_1)+V(\phi_2)]}/{\kB T}\}$ when equilibration is possible. 

Equation~(\ref{eq:FP-def}) belongs to the general class of stochastic systems with multiplicative noise, which implies that building a Langevin dynamics for the system requires special considerations. We first introduce the square-root of the mobility tensor $\Sigma$ via $M_{\alpha \beta}=\Sigma_{\alpha \nu} \Sigma_{\beta \nu}$, and then construct the corresponding Langevin equations [associated with Eq.~(\ref{eq:FP-def})] as 
\begin{eqnarray}
\dot{\phi}_\alpha(t)=M_{\alpha \beta}\left[-\mu_\phi V'(\phi_\beta)\right]+k_\mathrm{B}T\mu_\phi \, \Sigma_{\alpha \nu} \partial_\beta \Sigma_{\beta \nu} +\sqrt{2 k_\mathrm{B}T \mu_\phi} \,\Sigma_{\alpha \beta} \,\xi_\beta(t) ,
\label{eq:phasedynamics-noisy}
\end{eqnarray}
where $\xi$ satisfies $\langle \xi_\alpha(t) \xi_\beta(t') \rangle = \delta_{\alpha \beta} \delta(t-t')$. In Eq.~(\ref{eq:phasedynamics-noisy}), which coincides with Eq.~(2) in the main text, the Stratonovich convention for the multiplicative noise is used and a corresponding spurious drift term is introduced. This completes the stochastic formulation of the system of two mechanically coupled enzymes.

	The square-root $\Sigma$ of the mobility tensor $M$ can be calculated explicitly in the two limiting cases for $M$ described in the previous section. When  the phase constitutes the bottleneck of the dynamics and the mobility matrix is given by (\ref{eq:mob1b}--\ref{eq:mob12b}), we find $\Sigma_{11} = \Sigma_{22} = \nu_+ + \nu_-$ and $\Sigma_{12} = \Sigma_{21} = \nu_+ - \nu_-$, where we have defined $\nu_\pm \equiv (1/2)\sqrt{1 \pm h A_1 A_2 / (1-h^2)}$. When $h$ is small and the mobility matrix is given by (\ref{eq:mob1c}--\ref{eq:mob12c}) we find, again to first order in $h$, $\Sigma_{11}=1/\sqrt{1+A_1^2}$, $\Sigma_{22}=1/\sqrt{1+A_2^2}$, $\Sigma_{12}=\Sigma_{21}= h A_1 A_2 / \big[\sqrt{1+A_1^2} \sqrt{1+A_2^2} \big(\sqrt{1+A_1^2} + \sqrt{1+A_2^2} \big)\big]$.

\section*{Supplemental Figures}

\begin{figure}[h]
	\begin{center}
		\includegraphics[width=0.8\linewidth]{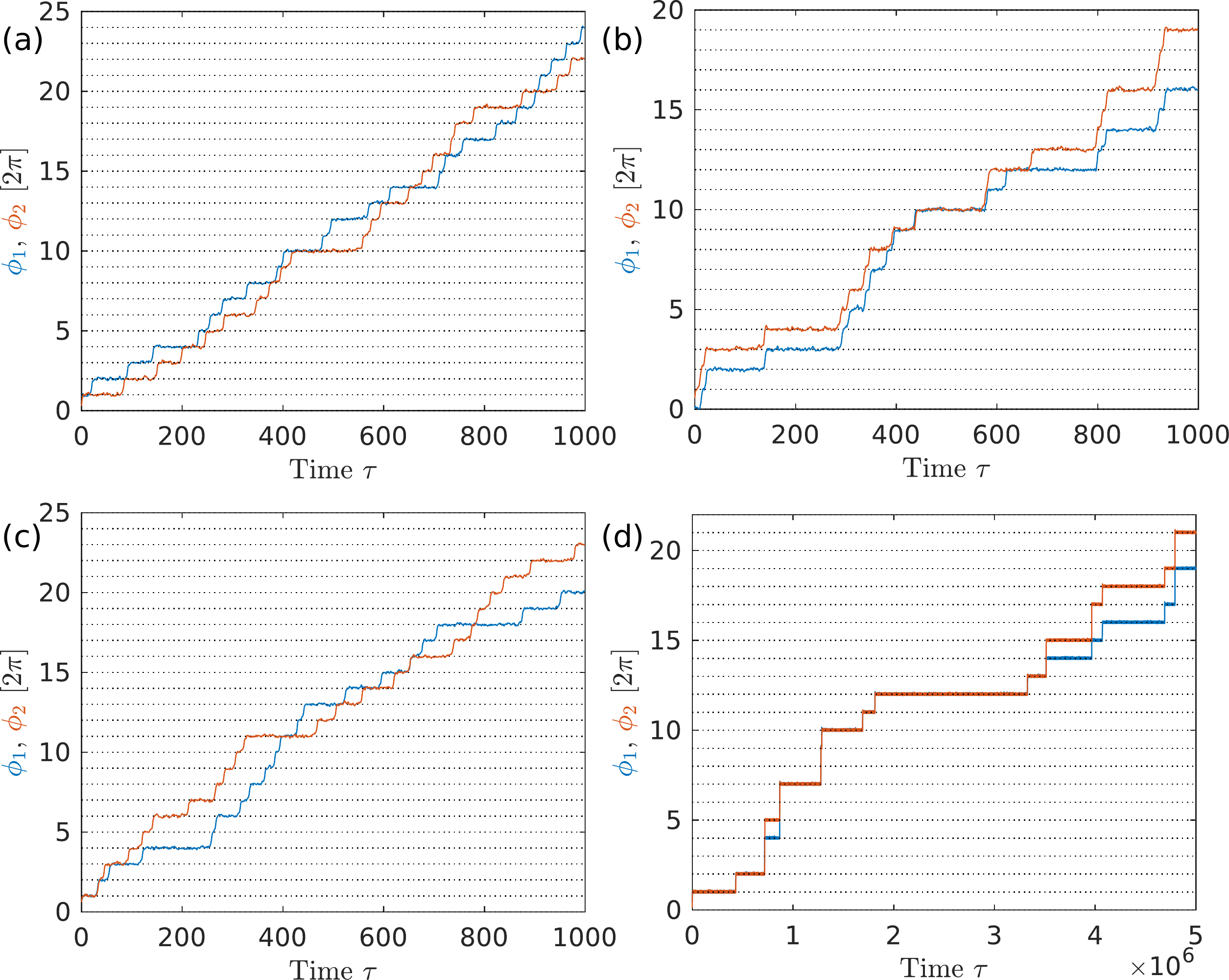}
		\caption{ Examples of time traces $\phi_{1,2}(t)$ corresponding to the same four systems  shown in Fig.~2(a) of the main text, showing how the enzyme steps are quasi-discrete, and stochastic in time. (a) Uncoupled system with $\bar{h}=0$, $\kB T/E_\mathrm{ba}=1$. (b) Synchronized system with $\bar{h}=-0.8$, $\kB T/E_\mathrm{ba}=1$. (c) Anticorrelated system with $\bar{h}=0.3$, $\kB T/E_\mathrm{ba}=1$. (d) Synchronized system at low noise with $\bar{h}=-0.8$, $\kB T/E_\mathrm{ba}=0.1$, showing clear multi-step runs. In all cases, $E_\mathrm{ba}/E_*=10^{-2}$. \label{fig:timetraces} }
	\end{center}
\end{figure}

\begin{figure}[h]
	\begin{center}
		\includegraphics[width=0.8\linewidth]{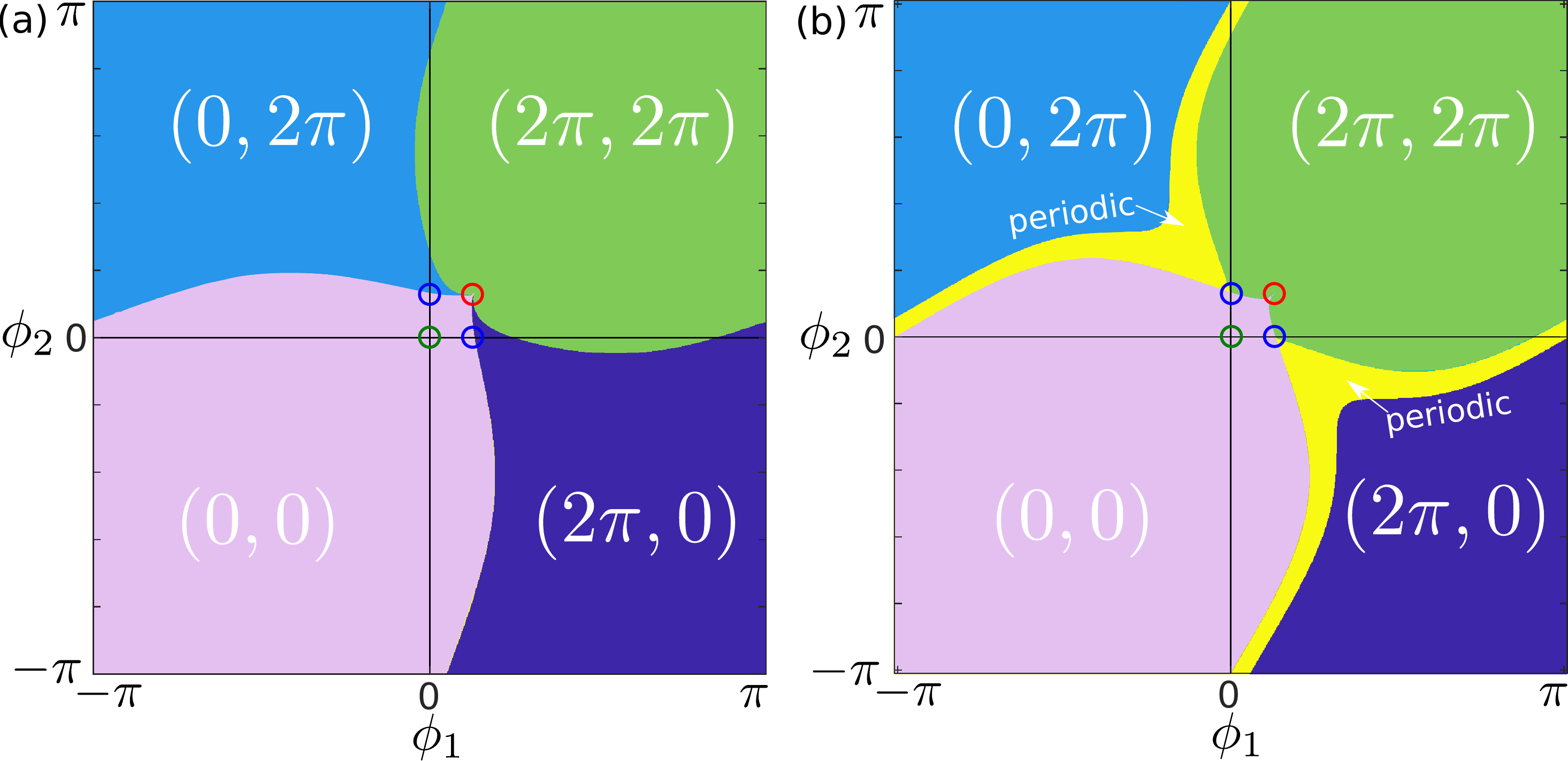}
		\caption{ The same bifurcation as described in the main text occurs when neither the conformational changes nor the phase changes constitute a bottleneck. Here, we have set $\mu_\phi \ell^2/\mu=1$ and used the full mobility matrix as given by (\ref{eq:mob1}--\ref{eq:mob12}). As the coupling is increased from (a) $h=-0.6$ to (b) $h=-0.8$, the bifurcation is crossed and a running band of periodic orbits emerges. The bias of the free energy landscape is set to $E_\mathrm{ba}/E_*=10^{-3}$. }
	\end{center}
\end{figure}

\begin{figure}[h]
	\begin{center}
		\includegraphics[width=0.8\linewidth]{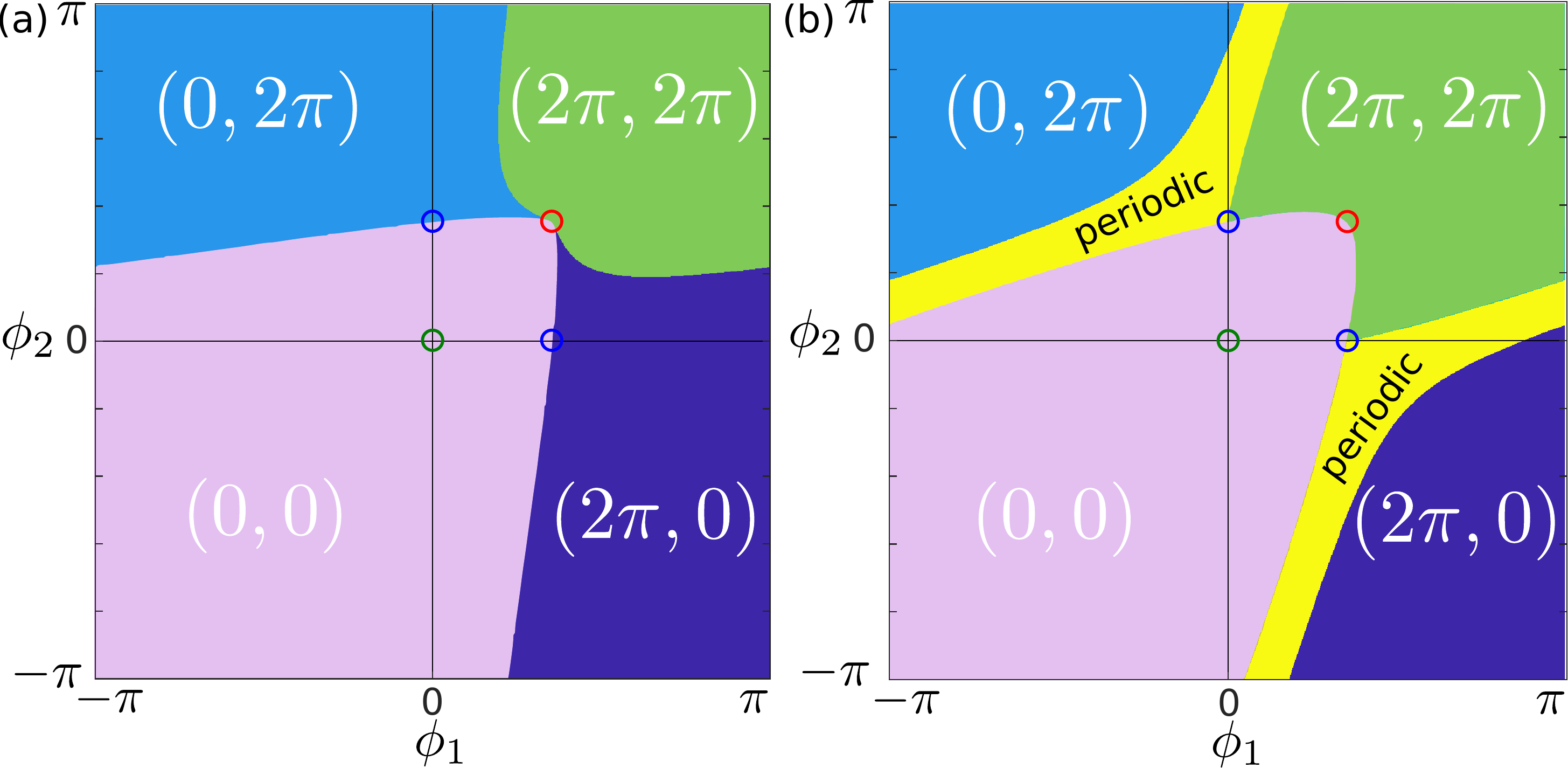}
		\caption{ The same bifurcation as described in the main text occurs in a simpler model, given by Eq.~(\ref{eq:fullphi}) but with a constant mobility matrix $M_{11}=M_{22}=1$ and $M_{12}=M_{21}=g$. As the coupling is increased from (a) $g=0.2$ to (b) $g=0.4$, the bifurcation is crossed and a running band of periodic orbits emerges. The bias of the free energy landscape is set to $E_\mathrm{ba}/E_*=0.02$.  }
	\end{center}
\end{figure}